\documentclass[aps,showpacs,preprintnumbers,superscriptaddress,twocolumn,floatfix]{revtex4-1}

\usepackage{amsmath,amssymb}
\usepackage{graphicx,color}

\def\be{\begin{equation}}
\def\ee{\end{equation}}
\def\bea{\begin{eqnarray}}
\def\eea{\end{eqnarray}}

\begin{document}

\title{Does space-time torsion determine the minimum mass of gravitating particles?}
\author{Christian G. B\"ohmer}
\email{c.boehmer@ucl.ac.uk}
\affiliation{Department of Mathematics, University College London, Gower Street, London
WC1E 6BT, United Kingdom}
\author{Piyabut Burikham}
\email{piyabut@gmail.com}
\affiliation{High Energy Physics Theory Group, Department of Physics, Faculty of Science,
Chulalongkorn University, Phyathai Rd., Bangkok 10330, Thailand}
\author{Tiberiu Harko}
\email{t.harko@ucl.ac.uk}
\affiliation{Department of Physics, Babes-Bolyai University, Kogalniceanu Street,
Cluj-Napoca 400084, Romania}
\affiliation{School of Physics, Sun Yat-Sen University, Guangzhou 510275, People's Republic of China}
\affiliation{Department of Mathematics, University College London, Gower Street, London
WC1E 6BT, United Kingdom}
\author{Matthew J. Lake}
\email{matthewj@nu.ac.th}
\affiliation{School of Physics, Sun Yat-Sen University, Guangzhou 510275, People's Republic of China}
\affiliation{The Institute for Fundamental Study, ``The Tah Poe Academia Institute", \\
Naresuan University, Phitsanulok 65000, Thailand}
\affiliation{Thailand Center of Excellence in Physics, Ministry of Education, Bangkok
10400, Thailand}

\date{\today }

\begin{abstract}
We derive upper and lower limits for the mass-radius ratio of spin-fluid spheres in Einstein-Cartan theory, with matter satisfying a linear barotropic equation of state, and in the presence of a cosmological
constant. Adopting a spherically symmetric interior geometry, we obtain the generalized continuity and Tolman-Oppenheimer-Volkoff equations for a Weyssenhoff spin-fluid in hydrostatic equilibrium, expressed in terms of the effective mass, density and pressure, all of which contain additional contributions from the spin. The generalized Buchdahl inequality, which remains valid at any point in the interior, is obtained, and general theoretical limits for the maximum and minimum mass-radius ratios are derived. As an application of our results we obtain gravitational red shift bounds for compact spin-fluid objects, which may (in principle) be used for observational tests of Einstein-Cartan theory in an astrophysical context. We also briefly consider applications of the torsion-induced minimum mass to the spin-generalized strong gravity model for baryons/mesons, and show that the existence of quantum spin imposes a lower bound for spinning particles, which almost exactly reproduces the electron mass.
\end{abstract}

\maketitle

\tableofcontents

\section{Introduction}
\label{sect1}

In a series of papers published around one hundred years ago, Cartan proposed an extension of Einstein's theory of general relativity in which the spin properties of matter act as an additional source for the gravitational field, influencing the geometry of space-time~\cite{Cartan}. In standard general relativity, space-time is described by a four-dimensional Riemannian  manifold $V_4$, and its source of curvature is assumed to be the energy-momentum tensor of the matter content. In~\cite{Cartan}, Cartan generalized Riemannian geometry by introducing connections with torsion, as well as an extended rule of parallel transport, referred to today as the Cartan displacement. From a mathematical point of view, torsion and the Cartan displacement are deeply related to the group of affine transformations, representing a generalization of the linear group of translations.

In Einstein-Cartan theory, matter sources space-time curvature as in general relativity. In addition, spin is postulated as the source of torsion in the Riemann-Cartan space-time manifold $U_4$~\cite{spin-tor}.
It is interesting to note that the concept of spin was introduced into theories of gravity, even before it was introduced into quantum mechanics, by Uhlenbeck and Goudsmit in 1925~\cite{Uhl}. Perfect fluids with spin were first studied by Weyssenhoff and Raabe~\cite{Weyssenhoff:1947} and are commonly referred to as Weyssenhoff fluids. (See~\cite{Boe1} for a detailed discussion of their physical geometric properties.) Later, an important development in the application of Einstein-Cartan gravity was the proposal by Kopczynski~\cite{Kop1} and Trautman~\cite{Traut} that the spin contributions of a Weyssenhoff fluid may avert the initial singularity at the Big Bang, by stopping the collapse in closed cosmological models at a minimum radius $R_{\rm min} \simeq 1$ cm. In Einstein-Cartan theory, this corresponds to a matter density $\rho _{\rm max} \simeq 10^{55}$ g/cm$^3$, so that, in the case of a chaotic spin distribution,
\begin{equation}
R_{\rm min}=\left(\frac{3G\hbar ^2n}{8mc^4}\right)^{1/3} \, , \quad \rho _{\rm max}=\frac{4m^2c^2}{3\pi ^2G\hbar ^2},
\end{equation}
where $n$ is the particle number density and $m$ is the mass of an individual particle with spin $\sim \hbar$.

It is important to note that, in Einstein-Cartan theory, {\it all} forms of rotation, including the angular momentum of an extended macroscopic body, a mass distribution of particles with randomly distributed spins, or an elementary particle with quantum mechanical spin, generate a modification of the standard Riemannian geometry of general relativity via torsion effects. However, in the following, we will adopt  the standard interpretation of Einstein-Cartan gravity, according to which the antisymmetric spin density of the theory is associated with the quantum mechanical spin of microscopic particles.

Thus, we use the term ``spinning fluid" to refer to a extended body whose infitesimal fluid elements possess nonzero orbital angular momentum density, derived from $SO(3)$ invariance, and the term ``spin-fluid" to refer to the course-grained (continuum) approximation of a large collection of particles, each possessing quantum mechanical spin. Hence, a spin-fluid may also be a {\it spinning} fluid, if it possess ``extrinsic" angular momentum in additional to ``intrinsic" $SU(2)$ spin. However, in the following, we will restrict our analysis to bodies with zero net orbital angular momentum, but a nontrivial intrinsic spin density.

At the macro-level, this approach yields a realistic a model of stable, static, compact astrophysical objects, composed of elementary quantum particles, while, on the micro-level, we take the continuum spin-fluid model at face value and apply it to the study of elementary particles themselves. In the latter, elementary ``point" particles are modeled as inherently extended bodies, and the resulting physical description qualitatively resembles that obtained in Dirac's extensible model of the electron \cite{Dirac_ext}.

In~\cite{Stew}, it was argued that the Big Bang singularity is only avoided due to the high degree of symmetry in the cosmological model used in~\cite{Kop1,Traut}. However, later studies demonstrated conclusively that, even in anisotropic cosmological models, the solutions of the Einstein-Cartan field equations do not lead to a singularity if the effect of torsion is greater than that of the shear~\cite{Kop2}. In~\cite{Gasp}, it was shown that early-epoch inflation may occur such that the dominant contributions to the effective energy-momentum tensor are given by the matter spin densities. A cosmic no-hair conjecture was also proven in Einstein-Cartan theory by taking into account the effects of spin in the matter fluid~\cite{H1}. If the ordinary matter forming the cosmological fluid satisfies the dominant and strong energy conditions, and the anisotropy energy $\sigma ^2$ is larger than the spin energy $S^2$, then all initially expanding Bianchi cosmologies - except Bianchi type IX - evolve toward the de Sitter space-time on a Hubble expansion time scale $\sim \sqrt{3/\Lambda}/c$. Static solutions of Einstein-Cartan theory with cylindrical and spherical symmetry were studied in~\cite{static}.

Realistic cosmological models in Einstein-Cartan theory were considered in~\cite{T2}, where it was shown that, by assuming the Frenkel condition~\cite{Frenkel}, the theory may be equivalently reformulated as an effective fluid model in standard general relativity, where the effective energy-momentum tensor contains additional spin-dependent terms. The dynamics of Weyssenhoff fluids were studied by Palle~\cite{Palle:1998} using  a $1+3$ covariant approach, and this approach was revised and extended in~\cite{Brechet:2007a,Brechet08}. An isotropic and homogeneous cosmological model in which dark energy is described by Weyssenhoff fluid, giving rise to the late-time accelerated expansion of the Universe, was proposed in~\cite{T3}, and observational constraints from Supernovae Type Ia were also discussed. These results show that, although the cosmological constant is still needed to explain current observations, the spin-fluid model contains some realistic features, and demonstrates that the presence of spin density in the cosmic fluid can influence the dynamics of the early Universe. Interestingly, for redshifts $z > 1$, it may be possible to observationally distinguish the spin-fluid model and the standard ``concordance" model of cold dark matter with a cosmological constant, assuming a spatially flat geometry.

In~\cite{T4} it was argued that, while spin-fluid dark energy models are statistically admissible from the point of view of the SNIa analysis, stricter limits obtained from Cosmic Microwave Background and Big Bang Nucleosynthesis constraints indicate that models with density parameters scaling as $a^{-6}(t)$ (a scaling that emerges naturally from a torsion dominated epoch) where $a(t)$ is the time-dependent scale factor of the Universe, are essentially ruled out by observations.  The effects of torsion in the framework of Einstein-Cartan theory in early-Universe cosmology were investigated in~\cite{T5}, while the gravitational collapse of a homogeneous Weyssenhoff fluid sphere, in the presence of a negative cosmological constant, was considered in~\cite{T6}. For recent investigations of the cosmology and astrophysics of Einstein-Cartan theory see~\cite{T7,T8,T9,T10,T11,T12,T13,T14,T15,T16}. In \cite{T17} it was shown  that by enlarging the Einstein–Cartan Lagrangian with suitable kinetic terms quadratic in the gravitational gauge field strengths (torsion and curvature) one can obtain some new, massive propagating gravitational degrees of freedom. It was also pointed out that this model has a close analogy to Fermi's effective four-fermion interaction and its emergent W and Z bosons.

In \cite{min1} it was shown that, within the framework of classical general relativity, the presence of a positive cosmological constant implies the existence of a minimum density in nature, such that
\begin{equation}
  \label{dens_lim}
  \frac{2GM}{c^2}\geq \frac{\Lambda }{6}R^3 \, , \quad
  \rho = \frac{3M}{4\pi R^3} \geq \rho_{\Lambda} \equiv \frac{\Lambda c^2}{16 \pi G} \, .
\end{equation}
These results follow rigorously from the generalized Buchdahl inequality for the Einstein-Hilbert action with an additional (positive) cosmological constant term ~\cite{min1}. The generalized Buchdahl inequality for charge-neutral, spherically symmetric, gravitating objects in the presence of $\Lambda > 0$ was first derived in~\cite{g1} and was shown to give rise to both maximum and minimum mass-radius ratios for stable compact objects. These results were further generalized to include the effects of charge~\cite{g2} and of an anisotropic interior pressure distribution~\cite{g3}. The effects of both charge and dark energy were considered in~\cite{m2}, yielding the lower mass-radius ratio bound
\begin{equation} \label{charged_min}
M \geq \frac{3}{4}\frac{Q^2}{Rc^2} + \frac{\Lambda R^3c^2}{12G} \, .
\end{equation}

Eqs.~(\ref{dens_lim}) and (\ref{charged_min})  give the lower bounds of the corresponding physical quantities only, and are consistent with previously obtained results in the limits $\Lambda \rightarrow 0$ and $Q \rightarrow 0$. Eq.~(\ref{dens_lim}) implies that $M \geq 0$, a result which is consistent with the Buchdahl limit \cite{Buch}, where no absolute lower bound can be established.

 Using an alternative approach, sharp bounds on the maximum mass-radius ratio for both neutral and charged, isotropic and anisotropic compact objects, in the presence of a cosmological constant, were rigorously derived in~\cite{sharp}. For fluids with isotropic pressure distributions and zero net charge ($Q=0$), in the absence of dark energy ($\Lambda = 0$), the maximum mass-radius ratio bound in all studies reduces to the classic result by Buchdahl, $2GM/(c^2R) \leq 8/9$~\cite{Buch}.  The Buchdahl compactness limit for a pure Lovelock static fluid star was obtained in \cite{Chak1}, where it was shown that the limit follows from the uniform density Schwarzschild's interior solution. For four-dimensional Einstein gravity, or for pure Lovelock gravity in $d=3N+1$ dimensions, Buchdahl's limit is equivalent to the criteria that gravitational field energy exterior to the star is less than half its gravitational mass. The Buchdahl bounds for a relativistic star in presence of the Kalb-Ramond field in four as well as in higher dimensions were derived in \cite{Chak2}.

Since a small but positive cosmological constant is still required in Einstein-Cartan theory, in order to explain late-time accelerated expansion~\cite{T3}, these results must be generalized to include the effects of spin (in the matter fluid) and torsion (in the space-time) in order to obtain realistic mass limits, either for fundamental particles or compact astrophysical objects. Though upper mass-radius ratio bounds are most relevant to the latter, lower mass-radius ratio limits may be applied, theoretically, to the former. In this case, one must ask the question: what is the gravitational radius of a fundamental particle?

For charged particles, Eq.~(\ref{charged_min}) gives rise to a classical minimum radius which, for $\Lambda R^2 \ll 1$, reduces approximately to the result obtained by Bekenstein, $R \gtrsim (3/4)Q^2/(Mc^2)$ ~\cite{Bekenstein:1971ej}. Essentially, this reproduces (up to a numerical factor of order unity) the classical radius of a charged particle, obtained by equating its rest mass with its electrostatic self-energy in special relativity. Hence, it may also be taken as a measure of the minimum classical gravitational radius of a charged particle in general relativity. Interestingly, this is also the length scale at which renormalization effects become important for charged particles in QED~\cite{QED}, suggesting a link between the gravitational and quantum mechanical theories.

Thus, in~\cite{Burikham:2015sro}, $R_{\rm min} \simeq Q^2/(Mc^2)$ was identified with the {\it total} minimum positional uncertainty $(\Delta x_{\rm total})_{\rm min}$, obtained by combining the canonical quantum uncertainty with gravitational/dark energy effects due to the existence of a finite horizon in space-times with $\Lambda >0$. In this model, a new form of minimum length uncertainty relation (MLUR), dubbed the ``dark energy uncertainty principle'" or DE-UP for short, which explicitly includes the de Sitter length $l_{\rm dS} = \sqrt{3/\Lambda}$ as well as the Planck length $l_{\rm Pl} = \sqrt{\hbar G/c^3}$, was proposed:
\begin{eqnarray} \label{gen}
\Delta x_{\rm total}  \geq  (\Delta x_{\rm canon.})_{\rm min} + \Delta x_{\rm grav}
\simeq \sqrt{\lambda_{\rm C}r} + \frac{l_{\rm Pl}^2l_{\rm dS}}{\lambda_{\rm C}r} \, .
\end{eqnarray}
Here, $(\Delta x_{\rm canon.})_{\rm min} \simeq \sqrt{\lambda_{\rm C}r}$ denotes the minimum possible canonical quantum uncertainty of a wave packet, corresponding to a particle with Compton wavelength $\lambda_{\rm C} = \hbar/(Mc)$, that has been freely evolving for a time $t = r/c$~\cite{Calmet:2004mp,Calmet:2005mh}. The term $\Delta x_{\rm grav}$ represents an additional contribution to the total uncertainty, due to the superposition of gravitational field states, which correspond to the superposition of position states associated with $M$. This, in turn, is equivalent to the uncertainty in the distance from $M$ to its horizon, $r_{\it H} \sim l_{\rm dS}$. Minimizing $\Delta x_{\rm total}$ with respect to either $M$ or $r$ and equating $R_{\rm min} \simeq (\Delta x_{\rm total})_{\rm min}$ yields $Q^2/(Mc^2) \lesssim (l_{\rm Pl}^2l_{\rm dS})^{1/3}$.

According to the model presented in~\cite{Burikham:2015sro}, this gives the maximum possible charge-squared to mass ratio for a stable, charged, self-gravitating and quantum mechanical object in general relativity with a positive cosmological constant. Assuming saturation of this bound for a particle that exists in nature and setting $Q^2 = e^2$ then gives
\be\label{me}
M \gtrsim \alpha_{e}(m_{\rm Pl}^2m_{\rm dS})^{1/3} = 7.332 \times 10^{-28} \, {\rm g},
\ee
where $m_{\rm Pl} = \sqrt{\hbar c/G}$ is the Planck mass, $m_{\rm dS} = \hbar/(cl_{\rm dS})$ is the ``de Sitter mass" and $\alpha_{\rm e} = e^2/(\hbar c) \simeq 1/137$ is the fine structure constant. The limiting value of $M$ is of the same of order of magnitude as the electron mass, $ m_{e} = 9.109 \times 10^{-28} \, {\rm g}$. Alternatively, rearranging the expression above gives
\begin{equation}  \label{Lambda_ident}
\Lambda \simeq \frac{\hbar^2G^2m_e^6c^6}{e^{12}} = \frac{l_{\rm Pl}^4}{r_{\rm e}^6} \simeq 10^{-56} \ {\rm cm^{-2}} \, ,
\end{equation}
where $r_{\rm e} = e^2/m_ec^2 \simeq 2.81 \times 10^{-13}$ cm is the classical electron radius. This representation of the cosmological constant in terms of the fundamental constants of nature is consistent with current observational constraints on the value of $\Lambda$~\cite{Planck}. Interestingly, an analogous formula derived in the context of strong gravity theory~\cite{strong} correctly predicts the order of magnitude value of the mass of the up quark, as the lightest known charged, quantum mechanical and strongly interacting particle~\cite{Burikham:2017bkn}.

Relation (\ref{Lambda_ident}) was first obtained by Nottale using a renormalization group approach~\cite{Nottale}, following work by Zel'dovich, who suggested that the dark energy density should be associated with the gravitational binding energy of electron-positron pairs spontaneously created in the vacuum~\cite{Zel'dovich:1968zz}. It was obtained independently in~\cite{min2}, with the use of Dirac's Large Number Hypothesis~\cite{LNH,Ray:2007cc} in the presence of a cosmological constant $\Lambda >0$, and in ~\cite{Beck:2008rd} using information theory considerations. A summary of the existing derivations of Eq.~(\ref{Lambda_ident}) is given in~\cite{Lake:2017ync}. We also note that the expression (\ref{Lambda_ident}) was used in~\cite{Wei} as the basis of a cosmological model in which $\Lambda \propto \alpha _e^{-6}$. From an observational perspective, it was shown in~\cite{Mariano}  that the value of the fine structure constant and the rate of the acceleration of the Universe  are better described by coinciding dipoles than by isotropic and homogeneous cosmological models.

For charge-neutral particles, the only ``available" radius is the Compton radius. Substituting $R = \lambda_{\rm C}$ into (\ref{dens_lim}) then gives
\bea \label{m_Lambda}
M \gtrsim m_{\Lambda} \equiv \sqrt{m_{\rm Pl}m_{\rm dS}} \simeq 10^{-35} {\rm g} ,
\eea
as the minimum mass of a stable, charge-neutral, gravitating and quantum mechanical object in general relativity with $\Lambda >0$~\cite{B1,Lake2017}. This is consistent with current experimental bounds on the mass of the electron neutrino obtained from Planck satellite data~\cite{Planck}. In addition, $m_{\Lambda} $ may be interpreted as the mass of an effective dark energy particle, associated with the Compton wavelength $l_{\Lambda} \equiv \sqrt{l_{\rm Pl}l_{\rm dS}}$, which is of the order of $0.1$ mm. According to this model, the dark energy density is approximately constant over large distances, but becomes granular on sub-millimetre scales, and it is notable that that tentative hints of periodic variation in the gravitational field strength on this length scale have recently been observed~\cite{submm}.

Upper and lower bounds on the mass-radius ratio for stable compact objects in extended gravity theories, in which modifications of the gravitational dynamics are described by a modified (effective) energy-momentum tensor, were obtained in~\cite{B1}, and their implications for holographic duality between bulk and boundary space-time degrees of freedom were investigated. The physical implications of the mass scale $M_T = (m_{\rm Pl}^2m_{\rm dS})^{1/3} \simeq m_{\rm e}/\alpha_{\rm e}$ were considered in~\cite{B4}, where, using the Generalized Uncertainty Principle (GUP)~\cite{GUP1}, it was shown that a black hole with age comparable to the age of the Universe may form a relic state with mass $M_T^{^{\prime}}= m_{\rm Pl}^2/M_{\rm T}$, rather than the Planck mass. The properties of the static AdS star were studied in~\cite{Burikham:2014ova}, where it was shown that, holographically, the universal mass limit corresponds to the upper limit of the deconfinement temperature in the dual gauge picture.

The brief summary above illustrates both the potential importance of spin and torsion in the gravitational dynamics of the Universe and, also, the fundamental importance of mass bounds for both macroscopic and microscopic objects. Such bounds have been derived for charged/uncharged, isotropic/anisotropic and classical/quantum objects, in the presence of dark energy and without. However, to date, most such bounds have been formulated within the context of general relativity or its analogues~\cite{Burikham:2017bkn}, or within a class of extended gravity theories which do not include torsion~\cite{B1}. Thus, it is the purpose of the present paper to consider the problem of upper and lower mass-radius ratio bounds for compact objects in Einstein-Cartan theory, in the presence of a cosmological constant. This represents a generalization of previous work to the important case of torsion gravity.

Thus, we obtain a spin-dependent generalization of the Buchdahl limit for the maximum mass-radius ratio of stable compact objects, which incorporates the effects of both torsion and dark energy, and we rigorously prove that a lower bound exists for spin-fluid objects, even in the absence of a cosmological constant. In the latter case, the lower limit is determined solely by the spin of the particles. In addition, we derive upper bounds on the physical and geometric parameters that characterize the spin-fluids using Ricci invariants. As a physical application of our results, we obtain absolute limits on the redshift for spin-fluid objects, which suggest that the observation of redshifts greater than two may indicate of the existence of space-time torsion. Hence, redshift observations can, at least in principle, detect the presence of torsion using compact objects. The implications of mass limits in a spin-generalized strong gravity theory, in which strong interactions and the properties of hadrons are investigated in a mathematical and physical framework analogous to Einstein-Cartan theory, are also briefly discussed. Bounds on the minimum mass of strongly interacting particles are obtained, and the role of spin in the mass relation is discussed.

This paper is organized as follows. The basic physical principles and mathematical formalism of Einstein-Cartan theory are briefly reviewed in Sec.~\ref{sect2}. In Sec.~\ref{sect3}, the gravitational field equations of Einstein-Cartan theory, in the presence of a cosmological constant, and for a static, spherically symmetric geometry are determined. The generalized Tolman-Oppenheimer-Volkoff equation is also obtained. The spin-generalized Buchdahl inequality, and maximum/minimum mass-radius ratio bounds for compact spin-fluid compact objects are derived in Section~\ref{sect4}, and complimentary bounds on the physical and geometric parameters obtained from the Ricci invariants are presented. Mass-radius ratio bounds in Einstein-Cartan theory with generic dark energy are derived in Sec.~\ref{sect5}. The astrophysical implications of our results are presented and discussed in Section~\ref{sect6}, where the upper limit for the gravitational redshift of compact objects is obtained. The implications of the lower mass-radius ratio bound for elementary particles are also discussed in the framework of an Einstein-Cartan type spin-generalized strong gravity theory. We briefly discuss and conclude our results in Section~\ref{sect7}.

\section{Einstein-Cartan theory and the Weyssenhoff fluid}
\label{sect2}

In the present Section we briefly review Einstein-Cartan theory and the inclusion of particle spin as a source of gravity. We also derive the gravitational field equations in a spherically symmetric geometry, obtain the generalized Tolman-Oppenheimer-Volkoff equation describing the hydrostatic equilibrium of a massive object, and discuss some specific models of the spin.

\subsection{Einstein-Cartan theory}

Einstein-Cartan theory is a \textit{geometric} extension of Einstein's theory of general relativity, which includes the spin-density of massive objects as a source of torsion in the space-time manifold.  The influence of the spin on the geometric properties and structure of space-time is thus a central feature of the theory, with fermionic fields such
as those of protons, neutrons and leptons providing natural sources of torsion \cite{spin-tor,Gasp, T2,T3,T4,T5,T6}. In standard general relativity, the source of curvature in the Riemannian space-time manifold $V_4$ is the matter energy-momentum tensor. In Einstein-Cartan theory, the Riemannian  space-time manifold is generalized to a Riemann-Cartan space-time manifold $U_4$, with nonzero torsion, and the spin of the matter fluid is assumed to act as its source \cite{spin-tor}. Thus, in the Einstein-Cartan theory, the spin-density tensor locally modifies the geometry of space-time, inducing a new geometric property, torsion.

In holonomic coordinates the torsion tensor ${T^{\lambda}}_{\mu\nu}$ is defined as the antisymmetric part of the affine connection ${{\tilde{\Gamma}}^{\lambda}}_{\ \mu\nu}$ \cite{spin-tor,Gasp,T2,T3,T4,T5,T6},
\begin{equation}  \label{torsion def}
{T^{\lambda}}_{\mu\nu}={\tilde{\Gamma}^{\lambda}}_{\ [\mu\nu]}= {\textstyle\frac{1}{2}}\left({\tilde{\Gamma}^{\lambda}}_{\ \mu\nu}-{\tilde{\Gamma}^{\lambda}}_{\ \nu\mu}\right) \, ,
\end{equation}
where a tilde denotes geometric objects in $U_4$ geometry 
In general relativity, the torsion tensor vanishes, due to the assumed symmetry of the connection in its two lower indices.

In Einstein-Cartan theory, the spin-connection 1-form $\tilde{\omega}^{\;\;\;\mu}_{\nu}$ can be split into two parts, a torsion free part (giving the usual spin-connection 1-form $\omega^{\mu}_{\;\;\;\nu}$, which is related to
the standard Christoffel symbol $\Gamma^{\alpha}_{\mu \nu}$) and a contortion 1-form $K^{\mu}_{\;\;\;\nu}$, which is related to the the torsion of space-time, so that \cite{spin-tor,Boe1}
\begin{equation}  \label{eq:cos6}
\tilde{\omega}^{\mu}_{\;\;\;\nu} = \omega^{\mu}_{\;\;\;\nu} + K^{\mu}_{\;\;\;\nu} \, .
\end{equation}
The torsion vector $T^{\mu}$  and the contortion tensor $K^{\mu}_{\;\;\;\nu}$ are related via \cite{spin-tor,Boe1}
\begin{equation}  \label{eq:cos6a}
T^{\mu} = {\rm D}e^{\mu} = {\rm d}e^{\mu} + \tilde{\omega}^{\mu}_{\;\;\;\nu} e^{\nu} = K^{\mu}_{\;\;\;\nu} \wedge e^{\nu} ,
\end{equation}
where we have used the fact that $\omega ^{\mu }_{\;\;\;\nu}$ is a torsion-free (Riemannian) connection. The above relation between torsion and contortion implies that their vector and axial vector components are related by
\begin{equation}  \label{eq:cos6b}
T_{[\mu \nu \lambda]} = K_{[\mu \nu \lambda]} \,  , \qquad T_{\mu \nu}^{\;\;\;\nu} = \frac{1}{2} K_{\nu \mu}^{\;\;\;\nu} .
\end{equation}

The gravitational field equations of Einstein-Cartan theory are derived by varying the usual Einstein-Hilbert action,
\be
S=\int{d^4x\sqrt{-g}\left[-\frac{\tilde{R}}{2k^2}+L_m\right]},
\ee
where $\kappa^2=8\pi G/c^4$ is the gravitational coupling constant, $\tilde{R}$ is the Einstein-Cartan curvature scalar constructed by using  the general asymmetric connection  ${{\tilde{\Gamma}}^{\lambda}}_{\ \mu\nu}$ of the $U_4$ manifold,
by taking the vielbein and the spin-connection as independent variables.  Hence the field equations in Einstein-Cartan theory can be written as \cite{spin-tor, Gasp}
\be  \label{eq:cos9}
R^{\mu}_{\;\;\;\nu} - \frac{1}{2}R \delta^{\mu}_{\nu} = \kappa ^2 \Sigma^{\mu}_{\;\;\;\nu}  ,
\ee
\be\label{eq:cos9b}
T^{\;\;\;\mu}_{\nu \lambda}+\delta^{\mu}_{\nu} T^{\;\;\;\alpha }_{\alpha \lambda}-\delta^{\mu}_{\lambda} T^{\;\;\;\alpha }_{\nu \alpha} = \kappa^2 s^{\;\;\;\mu }_{\nu \lambda} ,
\ee
where  $\Sigma^{\mu }_{\nu}$ denotes the canonical energy-momentum tensor, and $s^{\mu \nu \lambda} =\left(\delta L_m/\delta K_{\mu \nu \lambda }\right)/\sqrt{-g}$ is the canonical spin - density tensor of the matter fluid. Note that, here, we have implicitly included the existence of a cosmological constant term, which, for convenience, is incorporated on the right-hand side of the field equations in the definition of $\Sigma^{\mu}_{\nu}$. Where necessary, we will redefine the energy-momentum tensor such that $\Sigma^{\mu}_{\nu} \rightarrow \Sigma^{\mu}_{\nu} + \Lambda \delta ^{\mu}_{\nu}$, and in  the following analysis we will write the $\Lambda$-dependent terms explicitly. It is important to mention that
the equation governing the torsion tensor is an algebraic equation, and therefore the torsion is cannot
 propagate beyond the matter distribution, as, for example, a torsion wave. Hence {\it the torsion tensor does not vanish only inside material objects}. On the other hand, Einstein's field equations contain some additional terms that are quadratic in the torsion tensor \cite{spin-tor}.

\subsection{The Weyssenhoff spin-fluid}

We adopt the Weyssenhoff fluid model~\cite{Weyssenhoff:1947} for the description of matter with nonzero spin. From a physical point of view the Weyssenhoff fluid represents a continuous macroscopic medium (fluid), which is characterized on microscopic scales by the spin of the matter -- that is, by the individual spins of the particles which make up the ``fluid". The spin properties of the fluid, including the spin density, are described by an antisymmetric tensor $S_{\mu\nu}$~\cite{spin-tor,T2,T3, Brechet08} given by
\begin{align}
  \label{Sdt}
  S_{\mu\nu}=-S_{\nu\mu},
\end{align}
which is the source of the canonical spin - density tensor $s^{\lambda}{}_{\mu\nu}$ of the space-time, defined as
\begin{align}
  \label{tors}
  s^{\lambda}{}_{\mu\nu}=u^{\lambda}S_{\mu\nu},
\end{align}
where we have introduced the four-velocity of the fluid element $u^{\lambda }$. The Weyssenhoff spin-fluid also satisfies another important condition, the Frenkel condition~\cite{Frenkel}, which imposes the constraint that the intrinsic spin of the constituent particle of the fluid is space-like in the rest frame of the medium, so that
\begin{align}
  \label{Fr}
  S_{\mu\nu}u^{\nu}=0.
\end{align}
The Frenkel condition leads to an algebraic coupling between spin and
torsion, which can be written as
\begin{align}
  \label{ac}
  T^{\lambda}{}_{\mu\nu} = \kappa ^2 u^{\lambda}S_{\mu\nu}.
\end{align}
This follows from the fact that the torsion tensor becomes trace-free and hence the second and third terms on the left-hand side of (\ref{eq:cos9b}) vanish.

Mathematically, such a coupling also arises naturally when one performs the variation of the total action of the gravitational field--spinning fluid system~\cite{T2}. Thus, an important result in Einstein-Cartan theory is that the torsion contributions to the gravitational field equations are completely described by the spin density of the fluid. The spin-density scalar is an important and useful physical quantity, which is defined as ~\cite{spin-tor,T2,T3}
\begin{align}
  S^2 \equiv \frac{1}{2} S_{\mu\nu}S^{\mu\nu}\geq 0.
  \label{Spin density scalar}
\end{align}

From a computational point of view the field equations of Einstein-Cartan theory simplify considerably for a perfect fluid source, reducing to the effective general-relativistic field equations with additional spin-dependent terms, and a spin field equation, respectively~\cite{spin-tor,T2,T3,T4,T5,T6}. The gravitational field equations can be formulated in the $V_4$ Riemann geometry as
\begin{align}
  R_{\mu\nu}-\frac{1}{2}g_{\mu \nu}R + \Lambda g_{\mu\nu}= \kappa ^2\; {^S}\Sigma _{\mu\nu}.
  \label{Einstein ef eq}
\end{align}
Here $\Lambda$ is the cosmological constant and the effective energy-momentum tensor of the spin-fluid is
\begin{align}
  \label{Emt}
        {}^S\Sigma_{\mu\nu} &= (\rho_{\rm eff}c^2+p_{\rm eff}) u_{\mu}u_{\nu} - p_{\rm eff}g_{\mu\nu}
        \nonumber \\ &-
        2(g^{\rho\lambda}-u^{\rho}u^{\lambda})
        \nabla_{\rho}\negmedspace\left[u_{(\mu}S_{\nu)\lambda}\right].
\end{align}
The effective mass density $\rho_{\rm eff}$ and the effective pressure $p_{\rm eff}$
are given by,
\begin{align}
  \label{rhos}
  \rho_{\rm eff} &= \rho - \kappa^2 S^2 = \rho _S, \\
  \label{ps}
  p_{\rm eff} &= p - c^2\kappa^2 S^2 = p_S,
\end{align}
where we introduce the torsional quantities $\rho_S=\rho -\kappa^2 S^2$ and $p_S=p-c^2 \kappa^2 S^2$. In the presence of the cosmological constant the total energy density $\rho _{tot}$  and total pressure $p_{tot}$ becomes
\be
\rho _{tot}=\rho_{{\rm eff}}+\frac{\Lambda}{\kappa ^2}=\rho - \kappa^2 S^2+\frac{\Lambda}{c^2\kappa ^2},
\ee
\be
p_{tot}=p_{{\rm eff}}-\frac{\Lambda}{\kappa ^2}= p - c^2\kappa^2 S^2- \frac{\Lambda}{\kappa ^2}.
\ee

The spin field equation is given by,
\begin{align}
  \label{Ef spin field equations}
  \nabla_{\lambda}\left(u^{\lambda}S_{\mu\nu}\right) =2u^{\rho}u_{[\mu\vphantom%
  ]}\nabla_{|\lambda}\left(u^{\lambda}S_{\vphantom[\rho|\nu]}\right).
\end{align}
If we assume the Frenkel condition (\ref{Fr}), then the spin contribution to
the energy-momentum tensor can be reformulated as~\cite{Boe1}
\begin{align}
  \label{eq:wey4a}
  u^{\alpha} \nabla_{\beta} \left(u^{\beta} S_{\alpha \mu}\right) &=
  u^{\alpha} S_{\alpha \mu} \nabla_{\nu} u^{\nu} + u^{\alpha } u^{\beta } \nabla_{\beta} S_{\alpha \mu}
  \nonumber \\ &=
  u^{\alpha } u^{\beta } \nabla_{\beta} S_{\alpha \mu} =
  -(u^{\beta } \nabla_{\beta} u^{\alpha}) S_{\alpha \mu}
  \nonumber \\ &=
  -a^{\alpha} S_{\alpha \mu}.
\end{align}
In the third and fourth step of the derivation the Frenkel condition was necessary for simplifying the results, and we have introduced the acceleration of the fluid $a^{\mu}$, defined by $a^{\mu}=\left(u^{\nu} \nabla_{\nu}\right)u^{\mu}$. In the following analysis, we restrict our study to the case for which the acceleration vanishes, $a^{\mu}\equiv 0$. for the sake of simplicity, we also assume that the physical energy density and pressure of the matter satisfy the linear barotropic equation of state,
\begin{align}
  \label{eos}
  p=w\rho c^2,
\end{align}
where $0\leq w\leq 1$ is the equation of state parameter.

\section{Static spherically symmetric fluid spheres in the Einstein-Cartan theory}
\label{sect3}

In the present Section we write down the interior field equations for a
static spherically symmetric geometry in Einstein-Cartan gravity, and we
derive the Tolman-Oppenheimer-Volkoff equation, describing the hydrostatic
equilibrium properties of spin-fluid spheres. Some simple models of the
torsion field are also introduced.

\subsection{Field equations of spin-fluid spheres}

As a starting point in our analysis we assume that the interior line element
for a spin-fluid is spherically symmetric, so that
\begin{align}
ds^{2}=e^{\nu }c^2dt^{2}-e^{\lambda }dr^{2}-r^{2}\left( d\theta ^{2}+\sin
^{2}\theta d\varphi ^{2}\right),
\end{align}
where the metric tensor components $\nu $ and $\lambda$ are functions of the
radial coordinate $r$ only. The components of the matter energy-momentum
tensor are
\begin{align}
  {^S}\Sigma_{0}^{0}=\rho_{S},\qquad
  {^S}\Sigma_{1}^{1}={^S}\Sigma_{2}^{2}={^S}\Sigma_{3}^{3}=-p_{S}.
\end{align}

The field equations describing the interior of a static spin-fluid sphere in Einstein-Cartan theory then take the form
\begin{align}
  \label{f1}
  -e^{-\lambda}\left(\frac{1}{r^2}-\frac{\lambda ^{\prime }}{r}\right)+\frac{1}{r^2} - \Lambda &=
  c^2\kappa ^2\left(\rho -\kappa ^2S^2\right), \\
  \label{f2}
  -e^{-\lambda}\left(\frac{\nu ^{\prime }}{r}+\frac{1}{r^2}\right)+\frac{1}{r^2} - \Lambda &=
  -\kappa ^2\left(p-c^2\kappa ^2S^2\right), \\
  \label{f3}
  p^{\prime } =-\frac{1}{2}\left(\rho c^2+p\right)\nu ^{\prime} &+\kappa^2
  S^2\left(\nu ^{\prime }+2\frac{S^{\prime }}{S}\right),
\end{align}
where Eq.~(\ref{f3}) follows from the conservation of the effective energy-momentum tensor, $\nabla_{\mu}({^S}\Sigma^{\mu}_{\nu})=0$. Eq.~(\ref{f1}) can be immediately integrated to give
\begin{align}
  \label{m1}
  e^{-\lambda}=1-\frac{2Gm_{\rm eff}(r)}{c^2r}-\frac{\Lambda }{3}r^2,
\end{align}
where we have defined the effective mass inside radius $r$ as
\begin{align}
  m_{\rm eff}(r) &= 4\pi \int_0^r \rho_S\, \bar{r}^2d\bar{r}
  \nonumber \\ &=
  4\pi \int_0^r{\left[\rho(\bar{r}) -\kappa ^2S^2(\bar{r})\right]\bar{r}^2d\bar{r}}.
\end{align}
By substituting Eqs.~(\ref{f3}) and (\ref{m1}) into Eq.~(\ref{f2}) we obtain
the generalized Tolman-Oppenheimer-Volkoff equation, describing the
equilibrium of spin-fluid spheres in Einstein-Cartan theory as
\begin{widetext}
\begin{align}
\frac{dp}{dr}=-\frac{\left(\rho c^2+p-2c^2\kappa ^2S^2\right)\left\{\left(\kappa ^2/2\right)\left[p-c^2\kappa ^2S^2-(2/3)\left(\Lambda /\kappa ^2\right)\right]r^3+Gm_{\rm eff}/c^2\right\}}{r^2\left(1-2Gm_{\rm eff}/c^2r-\Lambda r^2/3\right)}+c^2\kappa ^2\frac{d}{dr}S^2.
\end{align}
\end{widetext}
We note that this equation can also be conveniently written using the quantities $\rho_S$ and $p_S$. It then simplifies to
  \bea\label{37}
 && \frac{dp_S}{dr} = -\frac{\left(\rho_S c^2+p_S\right)}{r^2\left(1-2Gm_{\rm eff}/c^2r-\Lambda r^2/3\right)}\times \nonumber\\
  &&\left\{\left(\kappa ^2/2\right)\left[p_S-(2/3)\left(\Lambda /\kappa ^2\right)\right]r^3+Gm_{\rm eff}/c^2\right\},
  \eea
  \be\label{38}
\hspace{-4.5cm} \frac{dm_{\rm eff}}{dr} = 4\pi r^2 \rho_S.
  \ee
Formally this system of equations cannot be distinguished from the corresponding equations in the absence of torsion.

\subsection{Models for the torsion}

In the present Section we will briefly review some of the physical and geometrical models proposed to describe torsion in the framework of Einstein-Cartan theory.

\subsubsection{The constant torsion model}

The simplest assumption one can make about the averaged microscopic spin density is that it has a constant value inside the fluid, so that $S^2=S_0^2=\mathrm{constant}$. This choice simplifies the field equations considerably. However, one is faced with a serious drawback. Due to the algebraic field equations for torsion, the vacuum region of space-time must be torsion-free. Therefore, a physically viable star should satisfy the condition of vanishing torsion at the surface, in additional to the vanishing pressure which, in general relativity, defines the vacuum boundary. This is the most conservative model one can build.

If one assumes that the ``vacuum" region contains some remnant torsion, for instance torsion on cosmological scales, then one could relax this condition and consider solutions where the torsion does not vanish at the boundary but instead takes, for example, the value of the cosmological background torsion.

\subsubsection{The general-relativistic conservation equation}

A second form of the spin scalar can be obtained by imposing the condition that the thermodynamic parameters of the spin-fluid still satisfy the standard general relativistic conservation equation $\nu' = -2p'/\left(\rho c^2+p\right)$, see~\cite{static}, which gives the radial spin variation equation
\begin{align}
  \nu^{\prime }+2\frac{S^{\prime }}{S}=0.
\end{align}
In turn, this fixes the spin dependence of the metric as
\begin{align}
  S^2=S_0^2e^{-\nu}=S_0^2e^{\int{\frac{2dp}{\rho c^2+p}}}=S_0^2\rho^{2w/(1+w)},
  \label{spin_dep}
\end{align}
where $S_0$ is an arbitrary constant of integration, and we have used the linear barotropic equation of state (\ref{eos}).

As in the previous case, this poses serious problems to the theory. For linear and polytropic equations of state, the vanishing pressure surface coincides with the vanishing density surface. This means there exists some radius $R$ where $\rho(R)=p(R)=0$. Then (\ref{spin_dep}) implies $S(R)=0$ which appears consistent. However, the problematic point is that $e^{-\nu(R)} = 0$. Therefore, the metric function $e^{\nu}$ becomes divergent and the boundary of the star. Consequently, solutions of this type are also not desirable.

\subsubsection{The Fermion model}

A similar dependence of the spin on the energy density can be obtained as follows~\cite{static,Gasp}. We assume that the compact object consists of an ideal fluid made of fermions and that there is no overall polarisation of the spins. It was shown in~\cite{Gasp} that the contribution to the energy-momentum tensor then takes the form
\begin{align}
  S^2 =\frac{1}{8}\hbar ^2\left<n^2\right>=\frac{1}{8}\hbar^2 A_w^{-2/(1+w)} \rho^{2/(1+w)}, \label{fermtor}
\end{align}
where the matter is assumed to satisfy the linear barotropic equation of state (\ref{eos}). Here, $A_w$ is a dimensional constant depending on the parameter $w$ of the equation of state.

The functional form of this torsion contribution is similar to Eq.~(\ref{spin_dep}), but is without any link to the metric functions. Consequently, this model is the most viable physical model discussed so far.

\subsection{Constant density stars in Einstein-Cartan theory} \label{sect3.3}

Constant density stars, with $\rho =\rho _{0}=\mathrm{const.}$ are important toy models for estimating general relativistic/modified gravity effects on stellar properties. In the following, we briefly investigate the properties of constant density stars in Einstein-Cartan theory. For simplicity we assume first that the cosmological constant can be neglected, setting $\Lambda =0$ in the following analysis. In order to close the system of equations, we consider a \textit{pressure dependent} ``equation of state" for the torsion
\begin{align}
  S^{2}=\beta p^{\ell},
\end{align}
where $\beta $ and $\ell$ are constants. The variation of the effective mass and thermodynamic pressure, as functions of the radial coordinate $r$, are then described by
\begin{align}
  \frac{dm_{\rm eff}(r)}{dr}=4\pi \left[ \rho _{0}-\kappa ^{2}\beta p^{l}\right] r^{2} ,
  \label{const1}
\end{align}
together with the corresponding TOV equation
\begin{widetext}
\begin{align}
  \label{const2}
  \frac{dp}{dr}= -
  \frac{
    \left(\rho _{0}c^{2}+p-2c^{2}\kappa ^{2}\beta p^{\ell}\right)
    \left\{ \left( \kappa^{2}/2\right) \left( p-c^{2}\kappa ^{2}\beta p^{\ell}\right)r^3+Gm_{\rm eff}/c^{2}\right\}
  }
       {r^{2}\left( 1-2Gm_{\rm eff}/c^{2}r\right)\left( 1-c^{2}\kappa ^{2}\beta \ell p^{\ell-1}\right) }.
\end{align}
\end{widetext}
The system of Eqs.~(\ref{const1}) and (\ref{const2}) must be integrated subject to the boundary conditions $m_{\rm eff}(0)=0$, $p(0)=p_{c}$ and $p(R)=0$, where $R$ is the radius of the star and $p_{c}$ is the central pressure. By introducing a set of dimensionless variables $\left(\eta, M_{e\rm ff}, P\right) $, defined according to
\begin{align}
  r & =\frac{c}{\sqrt{4\pi G\rho _{0}}}\eta =10.362\times \left(\frac{\rho _0}{10^{15}\;{\rm g/cm^3}}\right)^{-1/2}\times \eta \;{\rm km},
  \nonumber \\
  m_{\rm eff} &= \frac{c^{3}}{\sqrt{4\pi G^{3}\rho _{0}}}M_{\rm eff} =\nonumber\\
  & 6.999\times \left(\frac{\rho _0}{10^{15}\;{\rm g/cm^3}}\right)^{-1/2} \times
  M_{\rm eff}\; \;M_{\odot} ,
  \label{const3}
\end{align}
and $p=\rho _{0}c^{2}P$, and by denoting
\begin{equation}
  \label{const4}
  B_l(l)=c^{2}\kappa ^{2}\beta \left( \rho _{0}c^{2}\right) ^{l-1},
\end{equation}
the structure equations (\ref{const1})-(\ref{const2}) can be rewritten in dimensionless form as
\be\label{const5}
\frac{dM_{\rm eff}(\eta)}{d\eta}=\left[1-B_l(l)P^l(\eta)\right]\eta^2 \, ,
\ee
and
\begin{widetext}
\be\label{const6}
\left[1-lB_l(l)P^{l-1}(\eta)\right]\frac{dP(\eta)}{d\eta}=-\frac{\left[1+P(\eta)-2B_l(l)P^l(\eta)\right]\left\{P(\eta)\left[1-B_l(l)P^{l-1}(\eta)\right]\eta ^3+M_{\rm eff}(\eta)\right\}}{\eta ^2\left(1-2M_{\rm eff}/\eta\right)} \, .
\ee
\end{widetext}
Eqs.~(\ref{const5}) and (\ref{const6}) must be integrated subject to the boundary conditions $M_{\rm eff}(0)=0$, $P(0)=p_c/\rho _0c^2$, and $P\left(\eta _S\right)=0$, where $\eta _S$ defines the vacuum boundary of the star. Hence, in order to obtain the boundary condition for the pressure at the center of the star, we need to fix the equation of state at $\eta =0$. In the following, we assume that {\it the central matter satisfies the Zeldovich, or ``stiff" equation of state}, so that $p_c=\rho _0c^2$. This choice fixes the central value of the dimensionless pressure as $P(0)=1$.

\begin{figure*}[!hbt]
\centering
\includegraphics[width=0.48\textwidth]{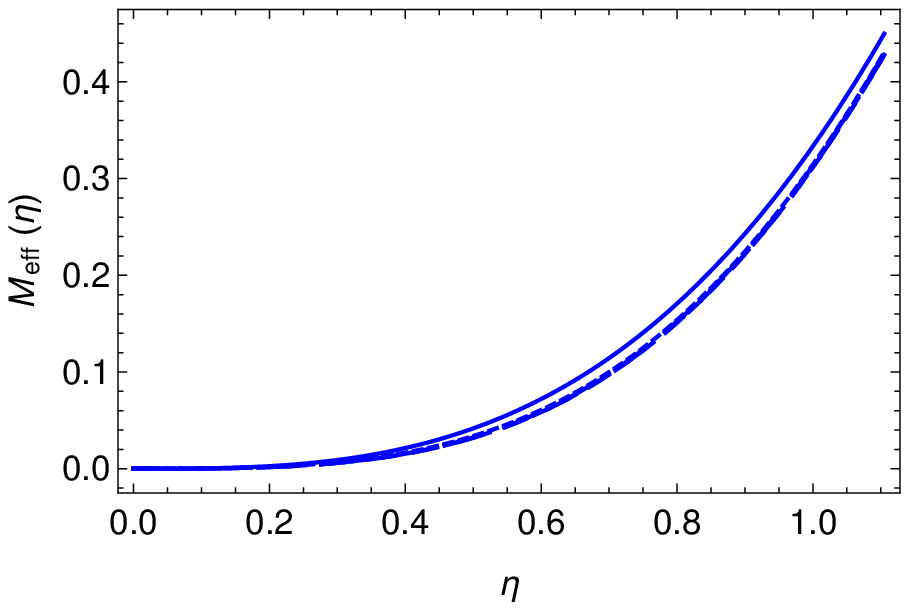}
\hfill
\includegraphics[width=0.48\textwidth]{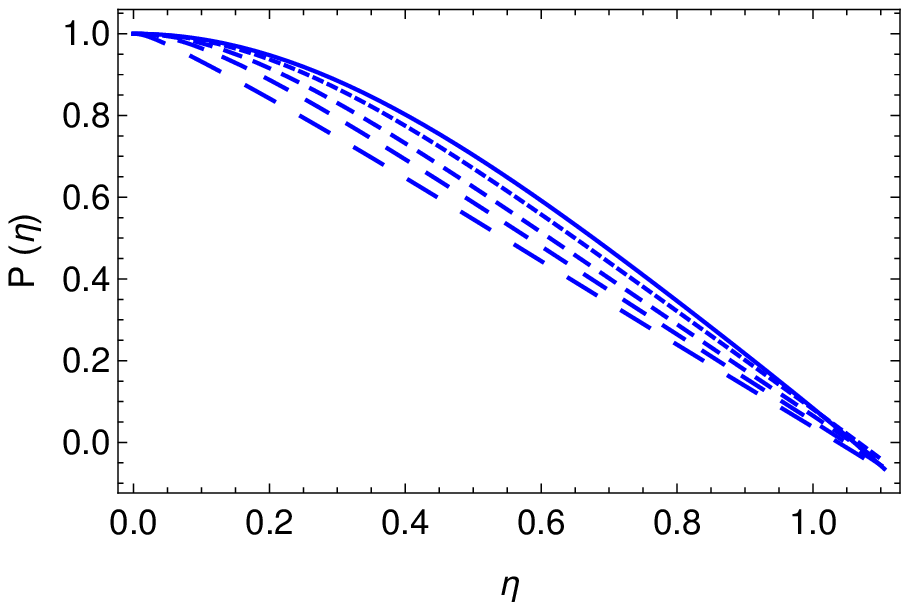}
\caption{Variation of the effective mass $M_{\rm eff}$ (left figure) and of the dimensionless pressure $P$ (right figure) as a function of the dimensionless radial coordinate $\eta$ for a star with spin density $S^2 \propto p^2$, for different values of the coefficient $B_2$: $B_2=0$ - the general relativistic limit - (solid curve), $B_2=0.3$ (dotted curve), $B_2=0.4$ (short dashed curve), $B_2=0.45$ (dashed curve) and $B_2=0.49$ (long dashed curve).}
\label{fig1}
\end{figure*}

In the following, for simplicity, we will consider only the case $l=2$, for which $S^2\propto p^2$.  We note that, for the choice $l=2$, the coefficient $B_2$ is given by $B_2=8\pi G\beta \rho _0=1.67\times 10^{9}\times \beta \times \left(\rho _0/10^{15}\;{\rm g/cm^3}\right)$. Hence, for the cases considered, the numerical values of $\beta $ are of the order of $\beta \approx 6\times 10^{-10}\times \left(\rho _0/10^{15}\;{\rm g/cm^3}\right)^{-1}$ s$^2$. The variation of the dimensionless mass $M_{\rm eff}$ and of the dimensionless pressure $P$ are presented, for different values of $B_2$, in Fig.~\ref{fig1}.

\begin{figure}[!hbt]
\centering
\includegraphics[width=0.48\textwidth]{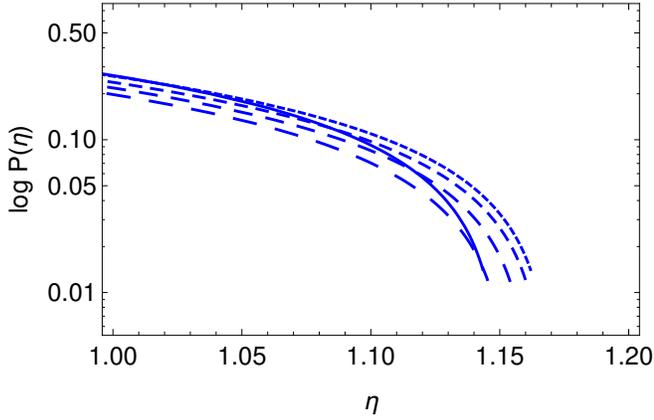}
\caption{Logarithmic plot of pressure for the same values of $B_2$ used in Fig. 1.}
\label{fig2}
\end{figure}

As one can see from the Figures, even in this simple case, the torsion has some small but observable effects on the global properties of compact astrophysical objects. The presence of torsion reduces the radius of the star from its general relativistic dimensionless radius $\eta_S\approx 1.06$ to a somewhat smaller value, $\eta _S\approx 1.02$. This value is not very sensitive to the assumed values of the parameter $B_2$. However, when looking at the behaviour of the solution near the vanishing pressure surface, some difference are clear, as one can see from Fig.~\ref{fig2}.

Hence, the radius of the star with the torsion effects taken into account is of the order of $R\approx 10.57\times \left(\rho _0/10^{15}\;{\rm g/cm^3}\right)^{-1/2}$ km, while for the standard general relativistic star we have $R\approx 10.98\times \left(\rho _0/10^{15}\;{\rm g/cm^3}\right)^{-1/2}$ km. This represents a discrepancy of less than 5\%. The same effect can be seen in the numerical values of the masses of the stars. While for the general relativistic case $M_{{\rm eff}}$ is of the order of $M_{{\rm eff}}\approx 0.38$, for stars in Einstein-Cartan theory $M_{{\rm eff}}$ has a slightly smaller value of order $M_{{\rm eff}}\approx 0.36$, which gives the corresponding masses values of order $M\approx 2.65\times \left(\rho _0/10^{15}\;{\rm g/cm^3}\right)^{-1/2}\;M_{\odot}$ and $M\approx 2.52\times \left(\rho _0/10^{15}\;{\rm g/cm^3}\right)^{-1/2}\;M_{\odot}$, respectively. This corresponds to roughly a 5\% change in the mass due to torrion. A good knowledge of the equation of state of dense neutron matter, associated with high precision astronomical observations, may therefore lead to the possibility of discriminating Einstein-Cartan theory from general relativity in the study of compact astrophysical objects.

\section{Buchdahl limits in Einstein-Cartan theory}
\label{sect4}

In this Section, we investigate the effects of the spin density of the matter fluid on the upper and lower mass limits, obtained via the generalized Buchdahl inequality in Einstein-Cartan theory. For a rapidly rotating object, the spherical symmetry is lost, and all physical/geometrical quantities show an explicit dependence on the angular coordinates.  However, this may not be (necessarily) true in the case of  particles carrying intrinsic quantum mechanical spin. Therefore, in the following, we will tentatively assume that the only effect of the spin and, hence, of the torsion of the space-time, is to modify the thermodynamic parameters of the matter fluid, so that they take the effective forms given by Eqs.~(\ref{rhos}) and (\ref{ps}), without influencing the spherical symmetry of the system. The upper and lower mass bounds can then be obtained in an analogous way to general relativity.

\subsection{The Buchdahl inequality in Einstein-Cartan theory}
\label{sect4.1}

The gravitational properties of a compact, static, spin-fluid sphere can be described in Einstein-Cartan theory by the spherically symmetric gravitational
structure equations, Eqs.~(\ref{37}) and (\ref{38}), respectively, and
\begin{equation}  \label{3a}
\frac{d\nu }{dr}=2\frac{ \left[ \frac{Gm_{\rm eff}}{c^2}+\frac{\kappa ^2}{2}\left( p_S-\frac{2\Lambda }{3\kappa ^2}\right) r^{3}\right] }{r^{2}\left(1-\frac{2Gm_{\rm eff}}{c^2r}-\frac{1 }{3}\Lambda r^{2}\right)} \, .
\end{equation}
To obtain Eq.~(\ref{3a}) we proceed as follows: we first solve Eq.~(\ref{f2}) for $\nu'$ then substitute $e^{-\lambda}$, as given by Eq.~(\ref{m1}), into the resulting expression.

The system of the stellar structure equations given by Eqs.~(\ref{37}), (\ref{38}) and (\ref{3a}) must be considered together with an equation of state for the spin-fluid, $p=p(\rho)$, and subject to the boundary conditions $p(R)=0 $, $p(0)=p_{c}$, $\rho (0)=\rho _{c}$, $S^2(0)=0$ and $S^2(R)=\sigma _0^2$, where $\rho_{c}$ and $p_{c} $ are the density and pressure at the centre of the sphere, respectively.

With the use of Eqs.~(\ref{37}), (\ref{38}), and (\ref{3a}), it is straightforward to show that the metric function $\zeta (r)=e^{\frac{\nu (r) }{2}}>0$, which is positive everywhere within the interior, $\zeta (r)>0$, for all $r \in \lbrack 0,R\rbrack$, satisfies the following differential equation~\cite{g1}:
\begin{widetext}
\bea\label{4}
\frac{1}{r}\sqrt{1-\frac{2Gm_{\rm eff}(r)}{c^2r}-\frac{1 }{3}\Lambda r^{2}}\frac{d}{dr}\left[ \sqrt{1-\frac{2Gm_{\rm eff}(r)}{c^2r}-\frac{1 }{3}\Lambda r^{2}}\frac{1}{r}\frac{d\zeta (r)}{dr}\right]
= \frac{\zeta (r)}{r}\frac{d}{dr}\frac{Gm_{\rm eff}(r)}{c^2r^{3}} \, .
\eea
\end{widetext}
This equation is formally analogous to its general-relativisitc counterpart, with effective spin-dependent quantities taking the place of standard thermodynamic variables.

As a next step in our analysis, we adopt the fundamental assumption that the effective density $\rho _{\rm eff}$ does {\it not} increase with increasing radial distance $r$. It therefore follows that the mean effective density of the matter distribution, $\langle \rho _{\rm eff} \rangle =3m_{\rm eff}(r)/4\pi r^{3}$, located inside radius $r$, does not increase either. Hence it follows that, as in standard general relativity, the condition
\begin{equation}
\frac{d}{dr}\frac{m_{\rm eff}(r)}{r^{3}}<0 \, ,
\end{equation}
must hold independently of the equation of state of the matter. This is a crucial assumption in the following analysis.  The simplest way to satisfy it is to assume that the spin density scalar $S^2$ itself is a monotonically decreasing function of the radial coordinate inside the star, reaching its maximum value for $r=0$. This assumption is also consistent with the requirement that the torsion takes vanishingly small values outside the vacuum boundary of the dense astrophysical object. On the other hand, a constant value of the torsion inside the compact object is also possible, with the condition that the effective energy density is always positive, and monotonically decreases. However, we note that there may be torsion models of stars that do not satisfy this assumption. Consequently, our results would not apply in such cases.

By introducing a new independent variable, defined as~\cite{g1}
\begin{equation}
\xi =\int_{0}^{r}r^{\prime }\left( 1-\frac{2Gm_{\rm eff}(r^{\prime })}{c^2r^{\prime }}-\frac{1 }{3}\Lambda r^{\prime 2}\right) ^{-\frac{1}{2}}dr^{\prime } \, ,
\end{equation}
we obtain, from Eq.(\ref{4}), the fundamental result that in Einstein-Cartan theory {\it all} stellar-type spin-fluid distributions with negative density gradient satisfy the condition
\begin{equation}  \label{5}
  \frac{d^{2}}{d\xi ^{2}} \left( e^{\frac{\nu \left( \xi \right) }{2}} \right)<0 \, , \quad \forall r\in \left[ 0,R\right] \, .
\end{equation}
With the use of the mean value theorem, it follows that
\begin{equation}
  \frac{d}{d\xi } \left(e^{\frac{\nu \left( \xi \right) }{2}}\right) \leq
  \frac{e^{\frac{\nu\left( \xi \right) }{2}}-e^{\frac{\nu \left( 0\right) }{2}}}{\xi } \, ,
\end{equation}
or, by taking into account that $e^{\frac{\nu \left( 0\right) }{2}}>0$, we obtain
\begin{equation}
  \frac{d}{d\xi } \left(e^{\frac{\nu \left( \xi \right) }{2}}\right) \leq \frac{e^{\frac{\nu \left( \xi \right) }{2}}}{\xi } \, .
\end{equation}
In terms of our initial variables $\left(m_{\rm eff},p_S,\Lambda \right)$ we therefore obtain the inequality
\begin{widetext}
\begin{align}\label{6}
\frac{(G/c^2)m_{\rm eff}(r) + (\kappa^2/2) \left[ p_S(r)-\frac{2\Lambda }{3\kappa^2}\right] r^{3}}{r^{3}\sqrt{1-\frac{2Gm_{\rm eff}}{c^2r}-\frac{1 }{3}\Lambda r^{2}}}\leq
\left[\int_{0}^{r}r^{\prime }\left( 1-\frac{2Gm_{\rm eff}(r^{\prime })}{c^2r^{\prime }}-\frac{1 }{3}\Lambda r^{\prime 2}\right) ^{-\frac{1}{2}}dr^{\prime }\right]^{-1} \, .
\end{align}
\end{widetext}

Since, as already pointed out, for stable compact objects the mean density $m_{\rm eff}/r^{3}$ does not increase outwards, it follows that
\begin{equation}
\frac{m_{\rm eff}(r^{\prime })}{r^{\prime }}\geq \frac{m_{\rm eff}(r)}{r}\left(\frac{r^{\prime }}{r}\right) ^{2} \, , \quad \forall r^{\prime }\leq r \, .
\end{equation}
For convenience, we now introduce the dimensionless variable $\alpha (r)$, defined as
\begin{equation}
\alpha \left( r\right) =1+\frac{c^2 }{6G}\Lambda \frac{r^{3}}{m_{\rm eff}(r)} \, .
\end{equation}
Moreover, we assume that in Einstein-Cartan theory, in the presence of a cosmological constant, the condition~\cite{g1}
\begin{equation}
\frac{\alpha \left( r^{\prime }\right) m_{\rm eff}\left( r^{\prime }\right) }{r^{\prime }}\geq \frac{\alpha \left( r\right) m_{\rm eff}\left( r\right) }{r}\left( \frac{r^{\prime }}{r}\right) ^{2} \, ,
\end{equation}
or, equivalently
\begin{eqnarray}  \label{7}
&&\left( 1+\frac{c^2 }{6G}\Lambda \frac{r^{\prime 3}}{m_{\rm eff}\left(r^{\prime }\right) }\right) \frac{m_{\rm eff}\left( r^{\prime }\right) }{r^{\prime }} \geq
\notag \\
&& \left( 1+\frac{c^2 }{6G}\Lambda \frac{r^{3}}{m_{\rm eff}\left( r\right) }\right) \frac{m_{\rm eff}\left( r\right) }{r}\left( \frac{r^{\prime }}{r}\right)^{2} \, , \quad \forall r\in [0,R] \, ,  \notag \\
\end{eqnarray}
holds inside any compact spin-fluid object. In fact, the validity of Eq.~(\ref{7}) is independent of the sign of the cosmological constant $\Lambda$ and is generally valid for all spin-fluid matter distributions with decreasing density profiles. Hence, we can evaluate the right-hand side of Eq.~(\ref{6}) in the following way:
\begin{widetext}
\begin{eqnarray}\label{8}
\int_{0}^{r}\frac{r^{\prime }}{\left( 1-\frac{2Gm_{\rm eff}\left( r^{\prime }\right)}{c^2r^{\prime }}-\frac{1 }{3}\Lambda r^{\prime 2}\right) ^{\frac{1}{2}}}dr^{\prime }&=&\int_{0}^{r}\frac{r^{\prime }}{\left( 1-\frac{2\alpha \left(
r^{\prime }\right) Gm_{\rm eff}\left( r^{\prime }\right) }{c^2r^{\prime }}\right) ^{\frac{1}{2}}}dr^{\prime }
\geq \int_{0}^{r}\frac{r^{\prime }}{\left[ 1-\frac{2\alpha \left( r\right)G m_{\rm eff}(r)}{c^2r}\left( \frac{r^{\prime }}{r}\right) ^{2}\right] ^{\frac{1}{2}}}dr^{\prime }
\nonumber\\
&=& \frac{c^2r^{3}}{2\alpha \left( r\right) Gm_{\rm eff}(r)}\left[ 1-\left( 1-\frac{2\alpha \left( r\right) Gm_{\rm eff}(r)}{c^2r}\right) ^{\frac{1}{2}}\right] \, .
\end{eqnarray}
\end{widetext}

Finally, with the use of Eq.~(\ref{6}), Eq.~ (\ref{8}) yields the Buchdahl inequality for compact gravitating spheres in Einstein-Cartan theory,
\begin{equation}\label{9}
\frac{\frac{Gm_{\rm eff}\left( r\right)}{c^2}+\frac{\kappa ^2}{2} \left( p_S-\frac{2\Lambda }{3\kappa ^2}\right) r^{3}}{\sqrt{1-\frac{2Gm_{\rm eff}(r)}{c^2r}-\frac{1 }{3}\Lambda r^{2}}}
\leq \frac{\frac{2Gm_{\rm eff}(r)}{c^2}\left(1+\frac{c^2 }{6G}\Lambda \frac{r^{3}}{m_{\rm eff}(r)}\right) }{1-\sqrt{1-\frac{2Gm_{\rm eff}(r)}{c^2r}-\frac{1 }{3}\Lambda r^{2}}} \, ,
\end{equation}
which is valid for all $r$ inside the compact object. We note that this result does not depend on the sign  of the cosmological constant term $\Lambda$.

\subsection{The maximum mass-radius ratio bound for spin-fluid spheres}
\label{sect4.2}

Let us consider first the case $\Lambda =0$ and $S^2=0$. By evaluating Eq.~(\ref{9}) at the vacuum boundary of the object $r=R$, we obtain
\begin{equation}
\frac{1}{\sqrt{1-\frac{2GM}{c^2R}}}\leq 2\left[ 1-\left( 1-\frac{2GM}{c^2R}\right) ^{\frac{1}{2}}\right] ^{-1} \, ,
\end{equation}
where $M=m(R)$ is the total mass of the star, leading to the well-known Buchdahl limit for the maximum mass of stable, zero-spin density compact objects~\cite{Buch},
\begin{equation}  \label{BB}
\frac{2GM}{c^2R}\leq \frac{8}{9} \, .
\end{equation}
For $\Lambda \neq 0$, $S^2 \neq 0$, Eq.~(\ref{9}) leads, instead, to the following upper limit for the mass-radius ratio of compact spin-fluid sphere:
\begin{widetext}
\begin{equation}\label{10}
\frac{2GM_{\rm eff}}{c^2R}\leq \left( 1-\frac{1 }{3}\Lambda R^{2}\right)
\left[ 1-\frac{1}{9}\frac{\left( 1+\frac{3p_S(R)}{\left<\rho _{\rm eff}(R)\right>c^2}-\frac{2\Lambda }{\kappa ^2\left<\rho _{\rm eff}(R)\right>c^2}\right) ^{2}}{\left(1-\frac{1}{3}\Lambda R^{2}\right)\left(1+\frac{p_S(R)}{\left<\rho _{\rm eff}(R)\right>c^2}\right)^2}\right] \, ,
\end{equation}
\end{widetext}
where $M_{\rm eff}=m_{\rm eff}(R)$ is the total mass of the object, and we have denoted $\left<\rho _{\rm eff}(R)\right>=3M_{\rm eff}/4\pi R^3$.

\subsection{The minimum mass-radius ratio bound for spin-fluid spheres -- ``particles"}

\label{sect4.3}

The Buchdahl inequality (\ref{9}) can be rewritten in the equivalent form
\begin{equation}  \label{58}
\sqrt{1-\frac{2GM_{\rm eff}}{c^2R}-\frac{1}{3}\Lambda R^2}\geq \frac{\frac{GM_{\rm eff}}{c^2R}+\frac{\kappa ^2}{2}p_S(R)R^2-\frac{1}{3}\Lambda R^2}{\frac{3GM_{\rm eff}}{c^2R}+\frac{\kappa ^2}{2}p_S(R)R^2} \, .
\end{equation}
By introducing a new variable $u$, defined as
\begin{equation}
u=\frac{GM_{\rm eff}}{c^2R}+\frac{1}{6}\Lambda R^2 \, ,
\end{equation}
Eq.~(\ref{58}) can be rewritten as
\begin{equation}  \label{60}
\sqrt{1-2u}\geq \frac{u+a}{3u+a} \, ,
\end{equation}
where we have defined
\begin{equation}
a = \frac{\kappa ^2}{2}p_S(R)R^2-\frac{1}{2}\Lambda R^2 \,.
\end{equation}
Squaring Eq.~(\ref{60}), we obtain for $u$ the condition
\begin{equation}  \label{61}
uf(u)=u\left[9 u^2+(6 a-4) u+(a-2) a \right] \leq 0 \, .
\end{equation}

Assuming $u\neq 0$, the equation $f(u)$ has two real roots, and condition (\ref{61}) can be reformulated as
\begin{equation}  \label{63}
u\left(u-u_1\right)\left(u-u_2\right) \leq 0 \, ,
\end{equation}
where
\begin{align}
  u_1 &= \frac{1}{9} \left(2-3 a-\sqrt{6 a+4}\right),
  \nonumber \\
  u_2 &= \frac{1}{9} \left(2-3a+\sqrt{6 a+4}\right),
\end{align}
or, approximately
\begin{equation}
u_1 \simeq -\frac{1}{2}a \, , \quad u_2 \simeq \frac{4}{9}-\frac{a}{6} \, .
\end{equation}
The relation (\ref{63}) is satisfied if $u\geq u_1$, $u \leq u_2$, or $u\leq u_1$ and $u\geq u_2$. However, the second set of conditions would contradict the upper Buchdahl limit. Therefore, from $u_1\geq -a/2$, we obtain the lower mass-radius ratio bound for compact spin-fluid spheres in Einstein-Cartan theory as
\begin{equation}
\frac{2GM_{\rm eff}}{c^2R}\geq \frac{1}{6}\Lambda R^2-\frac{\kappa ^2}{2}p_S(R)R^2 \, .
\end{equation}
By also assuming that the thermodynamic pressure vanishes at the surface of the object, $p(R)=0$, we obtain $p_S(R)=-c^2\kappa ^2 S^2(R)$. Under these conditions, the lower mass-radius ratio bound reduces to
\begin{equation}  \label{68*}
\frac{2GM_{\rm eff}}{c^2R}\geq \frac{1}{6}\Lambda R^2+\frac{1}{2}c^2\kappa ^4 S^2(R)R^2 \, .
\end{equation}
Hence, even in the absence of a cosmological constant, the existence of space-time torsion gives a lower bound on the possible mass-radius ratio, for particles in nature with nonzero spin. This corresponds to a minimum mass density given by
\begin{equation}  \label{68}
\frac{2GM_{\rm eff}}{c^2R}\geq\frac{1}{2}c^2\kappa ^4 S^2(R)R^2 \, , \quad \rho _{\rm min}\geq \frac{3}{2}\kappa ^2S^2(R) \, .
\end{equation}
Note that these results require the presence of some torsion remnant in the exterior spacetime, otherwise $S(R)=0$ and one recovers the GR results.

\subsection{Bounds on the physical parameters from the Ricci invariants}

\label{sect4.4}

The scalar invariants of the Riemann tensor give important physical and geometrical information regarding the properties of compact objects, since they provide fully coordinate invariant characterizations of some important properties of physical systems, including curvature singularities and the Petrov type of the Weyl tensor, etc.~\cite{MH}. Two such scalar invariants, which have been extensively used in the
literature, are the trace of the energy-momentum tensor
\begin{equation}
  r_0=T=T_{\mu}^{\mu} \, ,
\end{equation}
and the square of the Ricci tensor
\begin{equation}
  r_1= R^{\mu \nu} R_{\mu \nu} \,.
\end{equation}
In order to find general restrictions on the physical and geometrical quantities for fluid spheres in Einstein-Cartan theory, we now consider the behavior of these invariants. If, inside the compact object, the static line-element is regular at all points $r$, satisfying the conditions $e^{\nu (0)}=\mathrm{constant}\neq 0$ and $e^{\lambda (0)}=1$, then the Ricci invariants must also be non-singular functions throughout the spin-fluid distribution. Consequently, for a regular space-time, the invariants are non-vanishing at the origin $r=0$. The invariant $r_0={^s}\Sigma ={^s}\Sigma _{\mu }^{\mu}$ is given by
\begin{equation}
r_0={^s}\Sigma _{\mu }^{\mu}=\rho _{\rm eff}c^2-3p_{\rm eff}=\rho c^2-3p+2c^2\kappa^2S^2 \, .
\end{equation}
Assuming that $r_0$ is a monotonically decreasing function of $r$, so that $r_0(0)\geq r_0(R)$, and that both the matter energy density and pressure vanish at the vacuum boundary of the
sphere, we obtain the restriction
\begin{equation}  \label{70}
2c^2\kappa ^2\left[S^2(R)-S^2(0)\right]\leq \rho _cc^2-3p_c\geq 0 \, ,
\end{equation}
where $\rho _c=\rho (0)$ and $p_c=p(0)$. It is interesting to note that, if the matter at the center of the star satisfies the equation of state for radiation, $\rho _cc^2=3p_c$, the spin scalar of the compact object has
the same value at the center and at the vacuum boundary, $S^2(R)=S^2(0)$. This equations is consistent with no torsion anywhere, as one would expect.

Next, we consider the restrictions that can be obtained from the study of the invariant $r_1$. For constant density spin-fluid spheres, this takes the form
\bea
  r_1 &=& R_{\mu \nu}R^{\mu \nu}=\kappa^4 \left({^s}\Sigma _{\mu \nu}-\frac{1}{2}g_{\mu\nu}{^s}\Sigma+\frac{\Lambda }{\kappa ^2}g_{\mu \nu}\right)\times \nonumber\\
  && \left({^s}\Sigma ^{\mu \nu}-\frac{1}{2}g^{\mu \nu}{^s}\;\Sigma+\frac{\Lambda }{\kappa ^2}g^{\mu \nu}\right)=
  \nonumber \\
 && \kappa^4\left(\rho _{tot}^2c^4+3p_{tot}^2\right)=
  \left[\kappa^2\left(\rho -\kappa^2S^2\right)c^2+\Lambda \right]^2+
  \nonumber \\
  &&3\left[\kappa^2(p-c^2\kappa ^2S^2)- \Lambda\right]^2 .
\eea
Assuming again that $r_1$ is a decreasing function of the radial coordinate, so that $r_1(0)\geq r_1(R)$, and that the effect of the cosmological constant can be neglected, as compared to the effect of the spin, we obtain the following restriction for the surface spin density of a stable compact object in Einstein-Cartan theory:
\begin{equation}
S^2(R)\leq \frac{1}{2c^2\kappa ^2}\sqrt{\left[\rho _c-\kappa ^2S^2(0)\right]^2c^4+3\left[p_c-c^2\kappa ^2S^2(0)\right]^2} \, .
\end{equation}

\section{Mass-radius ratio bounds in Einstein-Cartan theory with generic dark energy}
\label{sect5}

In this section, we consider the upper and lower mass-radius ratio bounds for spherical object in the presence of dark energy with generic equation of state, $P_{0}=w_{0}\rho_{0}c^{2}$ where $P_{0}~(\rho_{0})$ denotes the dark energy pressure~(energy density) respectively. Note that $\Lambda = \kappa^{2}\rho_{0}$.

\subsection{Generic mass-radius ratio bounds in Einstein-Cartan theory}

Following the analysis presented in~\cite{B1}, with the replacements $\rho \to \rho_{S}$, $P \to p_{S}$, we obtain
\begin{equation}
\frac{dP_{e}}{dr}= -\frac{(\rho_{S}c^{2}+P_{e})\left[ \left( \frac{\kappa^{2}}{c^{2}}P_{e}-\frac{2\Lambda}{3}\right)r^{3}+\frac{\kappa^{2}m_{\rm eff}(r)}{\Omega_{2}}\right]}{2r^{2}\left[ 1-\frac{\kappa^{2}m_{\rm eff}(r)}{\Omega_{2}r}-\frac{\Lambda r^{2}}{3}\right]} \, ,
\end{equation}
where the effective pressure $P_{e}\equiv p_{S}+(1+w_{0})\rho_{0}c^{2}$.  At the surface of the sphere $r=R$, this gives the inequality
\begin{equation}
\frac{1}{2}\left( \frac{\kappa^{2}}{c^{2}}P_{e}-\frac{2}{3}\Lambda\right)R^{2}+\frac{\kappa^{2}M_{\rm eff}c^{2}}{2\Omega_{2}R}\leq e^{-\lambda/2}(1+e^{-\lambda/2}),
\end{equation}
where all quantities take the value at $R$. The mass-radius ratio is then bounded by
\begin{equation}
u_{-}\leq\frac{\kappa^{2}M_{\rm eff}c^{2}}{\Omega_{2}R}\leq u_{+} \, ,
\end{equation}
where
\begin{equation}
u_{\pm}=\frac{2}{9}\left( 2-\frac{3\kappa^{2}P_{e}R^{2}}{2c^{2}}\pm \sqrt{4+\frac{3\kappa^{2}P_{e}R^{2}}{c^{2}}-3\Lambda R^{2}} \right) \, .
\label{gbounds}
\end{equation}
For $p_{S}=0$, we then have $\kappa^{2}P_{e}=(1+w_{0})\Lambda c^{2}$ at the object's surface, which leads to the \textit{universal} bounds given in~\cite{B1}. A nontrivial minimum bound exists when either
\begin{align*}
  (1)\ \Lambda > 0, w_{0}< -2/3,\quad \text{or} \quad
  (2)\ \Lambda < 0, w_{0}>-2/3.
\end{align*}

For the case where only $P=0$, we have
\begin{equation}
P_{e}=-\kappa^{2}c^{2}S^{2}+(1+w_{0})\frac{\Lambda c^{2}}{\kappa^{2}} \, ,
\end{equation}
and the bounds become
\begin{eqnarray}
\frac{2GM_{\rm eff}}{Rc^{2}}\Big|_{\mathrm{max,min}}&=&\frac{4}{9}\Bigg[ 1+\frac{3}{4}R^{2}\left(\kappa^{4}S^{2}-(1+w_{0}\right)\Lambda)
\notag \\
&&\pm \sqrt{1+\frac{3}{4}R^{2}(w_{0}\Lambda - \kappa^{4}S^{2})}\Bigg] \, .
\label{maxb}
\end{eqnarray}
Interestingly, both maximum and minimum bounds exist only when the torsion is bounded from above by
\begin{equation}
\kappa^{4}S^{2}\leq w_{0}\Lambda+\frac{4}{3R^{2}} \, .  \label{maxb1}
\end{equation}

Generically, the minimum bound in Eq.~(\ref{gbounds}) exists only when
\begin{align}
  P_{e}R^{2} &> \frac{2c^{2}}{\kappa^{2}}\left( 1+\sqrt{1-\frac{\Lambda R^{2}}{3}} \right) \quad \text{or}
  \nonumber \\
  P_{e}R^{2} &< \frac{2c^{2}}{\kappa^{2}}\left( 1-\sqrt{1-\frac{\Lambda R^{2}}{3}} \right) \,.
\end{align}
In the specific case where the matter pressure vanishes at the surface, and assuming $|\Lambda| R^{2}\ll 1$, the condition for the nontrivial minimum bound to exist becomes
\begin{alignat}{2}
  \kappa^{4}S^{2} &< -\frac{1}{R^{2}}+\left(\frac{4}{3}+w_{0}\right)\Lambda \, , \quad
  &\mathrm{for}\ &S^{2}<0 \, ,
  \nonumber \\
  \kappa^{4}S^{2} &> \left(\frac{2}{3}+w_{0}\right)\Lambda \, ,  \quad
  &\mathrm{for}\ &S^{2}>0 \, .
\end{alignat}
The minimum bound can exist for both positive and negative $\Lambda$.

\subsection{Holographic implications of the maximum and minimum mass-radius ratio bounds}

In the bulk space-time, torsion contributes negative energy density and pressure for $S^{2}>0$ and vice versa. This is a unique characteristic of torsion which is different from both ordinary matter and dark energy. For nonzero $\Lambda$, the bulk space-time has an asymptotic boundary. For $\Lambda>0$ this is a cosmological horizon, the de Sitter horizon $\sim \sqrt{3/\Lambda}$, whereas for $\Lambda<0$ an asymptotically AdS boundary exists instead. The holographic implication of the maximum mass-radius bound is that the maximum information content of the bulk space is equal to the number of quantum gravity ``bits" (i.e., Planck-sized patches $\sim l_{\rm Pl}^2$) on the boundary.

In the asymptotically AdS case, the bulk gravity theory has a dual gauge theory description on the AdS boundary~(see, for example,~\cite{Aharony:1999ti} and references therein for a review of holographic duality and the AdS/CFT correspondence). It was found by Hawking and Page~\cite{Hawking:1982dh} that AdS space-time at finite temperature has a number of thermal phases distinguished by the existence and size of the black hole in the background. After the proposal of the AdS/CFT correspondence, Witten~\cite{Witten:1998zw} argued that these AdS phases correspond to the thermal phases of the gauge-theory (CFT) living on the AdS boundary and that the Hawking-Page phase transition between the thermal AdS and large-mass AdS black hole space-times corresponds to the deconfinement phase transition of the dual gauge theory. The dual gauge theory on the boundary will undergo a deconfinement phase transition when the temperature exceeds the Hawking-Page temperature of the AdS bulk.

Naturally, if thermal radiation in the AdS bulk sufficiently accumulates, gravitational collapse will occur and a black hole will be formed. Therefore, gravitational collapse in the AdS bulk holographically corresponds to the deconfinement phase transition of the dual gauge matter on the AdS boundary. Consequently, maximum mass bounds for static objects in the bulk inevitably correspond to the minimum possible deconfinement temperature on the boundary. It was found in~\cite{Burikham:2014ova} that there exists a {\it universal} upper mass limit for a fermionic star in AdS space, which corresponds to the universal maximum Hawking-Page transition temperature.

It is only in the asymptotically AdS space that the black hole has the lowest possible temperature corresponding to a certain critical mass. At slightly above the critical mass, thermal space-time prefers to have lower free energy if a black hole with that mass is formed at the same temperature~\cite{Hawking:1982dh}. This is the deconfinement phase transition of the dual gauge theory living on the AdS boundary. The critical size of the black hole in the AdS space where the Hawking-Page transition occurs is approximately $R\simeq R_{\mathrm{AdS}}=\sqrt{3/\Lambda}$. We can then use the maximum mass-radius ratio bound in Eq.~(\ref{maxb}) to calculate the mass and the corresponding Hawking temperature of this transition. By approaching from the small AdS black hole branch, we find
\begin{equation}
T_{\mathrm{bh}}\sim \frac{1}{M}\gtrsim \frac{9}{4}\frac{1}{R_{\mathrm{AdS}}}>\frac{9}{4}\sqrt{\frac{\Lambda}{3}} \, .
\end{equation}
Retrieving all constants, the transition temperature is approximately
\begin{equation}
T \sim \frac{\hbar c^{3}}{k_{B}G}\sqrt{\Lambda} \, ,
\end{equation}
which is consistent with the well-known Hawking-Page temperature.

When both the effects of torsion and $\Lambda$ are relatively small, Eqs.~(\ref{maxb}) can be approximated as
\begin{eqnarray}
\frac{2GM_{\rm eff}}{Rc^{2}}\Big|_{\mathrm{max}}&\simeq&\frac{8}{9}\Bigg[ 1+\frac{3}{16}R^{2}\kappa^{4}S^{2}-\frac{3}{16}(2+w_{0})\Lambda R^{2}\Bigg] \, ,
\notag \\
\frac{2GM_{\rm eff}}{Rc^{2}}\Big|_{\mathrm{min}}&\simeq&\frac{1}{2}R^{2}\kappa^{4}S^{2}-\frac{1}{6}(2+3w_{0})\Lambda R^{2} \, .  \label{demin}
\end{eqnarray}
Thus, torsion can reduce the deconfinement temperature of the dual gauge matter living on the boundary by fractions of $\kappa^{4}S^{2}R_{\mathrm{AdS}}^{2}\simeq\kappa^{4}S^{2}/\Lambda$. The minimum bound will increase due to positive torsion $S^{2}>0$. Holographically, the mass gap of the strongly coupled gauge theory will be enlarged since it is dual to the minimum mass~\cite{Burikham:2017bkn}, i.e. torsion widens the mass gap in the dual gauge theory.

Another crucial point we would like to emphasize in the negative $\Lambda$/asymptotically AdS case is that there is no maximum mass-radius ratio bound for (large) AdS black holes. In 3+1 dimensions, the mass to horizon radius ratio of an AdS black hole, given by~\cite{Burikham:2014ova}
\begin{equation}
\frac{M}{r_{h}}=\frac{c^{2}}{2G}\Big( 1+\frac{r^{2}_{\rm h}}{R^{2}_{\mathrm{AdS}}}\Big) \, ,
\end{equation}
is \textit{not} bounded from above and yet, remarkably, is still bounded from below according to
\begin{equation}
\left.\frac{M}{r_{h}}\right|_{\mathrm{min}}=\frac{c^{2}}{2G} \, .
\end{equation}
It should be noted that, for black hole in AdS space, even the minimum value of $M/r_{\rm h}$ exceeds the Buchdahl maximum mass-radius ratio bound of a \textit{non-black
hole} compact object, $2GM/rc^{2}|_{\rm max}=8/9$. Once a black hole is formed, the bounds on compact spherical objects are not applicable anymore. A black hole in asymptotically flat space has a fixed mass-radius ratio such that $2GM/c^{2}r_{h}=1$, while a black hole in AdS has $2GM/c^{2}r_{h}\geq 1$. Notably, a black hole in dS space has $2GM/c^{2}r_{h}\leq 1$.

From Eq.~(\ref{maxb1}) we see that, if $-\Lambda=\sqrt{3/R_{\mathrm{AdS}}}>0, S=0$, the maximum and minimum bounds for non-black hole compact objects exist if
\begin{equation}
\frac{R}{R_{\mathrm{AdS}}}\leq \sqrt{\frac{4}{9w_{0}}} \, .
\end{equation}
This is the condition for the formation of a small black hole in AdS space for $R=r_{h}$ and with $w_{0}\sim\mathcal{O}(1)$. We should therefore interpret this as meaning that the maximum and minimum mass-radius ratio bounds in space-time with $\Lambda < 0$ only exist for ``AdS-small'' objects with $R<R_{\mathrm{AdS}}$. The generalization of this result to the case of  nonzero torsion gives
\begin{equation}
\frac{R}{R_{\mathrm{AdS}}}\leq \sqrt{\frac{4}{3(3w_{0}+\kappa^{4}S^{2}R_{\mathrm{AdS}}^{2})}} \, ,
\end{equation}
for the AdS-small condition $S^{2}\geq 0$, $w_{0}\sim \mathcal{O}(1)$.

A remarkable consequence of the minimum mass-radius bound ratio induced by torsion is the statement that {\it a fermionic particle with Planck radius must have a mass larger than the Planck mass $m_{\rm Pl}=\sqrt{\hbar c/G}$}. This can be easily shown as follows. Using the minimum mass-radius ratio bound in Eqs.~(\ref{68}) or (\ref{demin}) and the fermionic spin density in (\ref{fermtor}), and substituting $R=R_{\rm Pl}=\hbar/m_{\rm Pl}c$, we simply obtain $M_{\rm eff}>9m_{\rm Pl}/8$. Thus, torsion provides alternative interpretation of the Planck mass as the minimum mass of the fermionic particle with Planck radius.

\section{Astrophysical and particle physics applications}
\label{sect6}

In the present Section we briefly consider some astrophysical and particle physics applications of the spin-fluid mass-radius ratio bounds obtained in Einstein-Cartan theory. In particular we will point out the effect that the torsion of the spin-fluid can have on the gravitational redshift of electromagnetic radiation emitted from the surface of compact stars. In addition, we will investigate the minimum mass-radius ratio bound in the framework of the strong gravity description of strong interactions, which offers an alternative (geometric) description of Yang-Mills type theories. In the latter case, we note that the strong gravity description is valid only approximately and as an {\it effective} theory for the gauge-singlet sector of QCD. Hence, it may be used as an effective theory to study confinement but {\it not} to describe scattering processes involving $SU(3)$ color charge (see \cite{Burikham:2017bkn} for further explanation).

\subsection{Gravitational redshift}

The existence of limiting values of the mass-radius ratio also leads to the existence of upper/lower bounds for other physical and geometrical quantities of observational interest. One important quantity is the surface red shift $z$, defined in the Schwarzschild-de Sitter geometry as~\cite{g1}
\begin{equation}  \label{z}
z=\left( 1-\frac{2GM_{\rm eff}}{c^2R}-\frac{1}{3}\Lambda R^{2}\right) ^{-\frac{1}{2}}-1 \, .
\end{equation}
In standard general relativity, and in the absence of the cosmological constant, the use of the Buchdahl bound (\ref{BB}) leads to the well-known constraint $z\leq 2$~\cite{g1}.

For spin-fluid compact objects in Einstein-Cartan theory, the surface red shift must obey the following restriction, which follows immediately from the generalized Buchdahl inequality (\ref{9}):
\begin{equation}
  \label{13}
  z \leq \frac{\frac{2GM_{\rm eff}}{c^2R}+\frac{1}{3}\Lambda R^2}{\frac{GM_{\rm eff}}{c^2R}+\frac{\kappa ^2}{2}\left[p_S(R)-\frac{2\Lambda }{3\kappa ^2}\right]R^2} \, .
\end{equation}
If the cosmological constant as well as the surface pressure vanish, $\Lambda \equiv 0$, $p(R)=0$, then the upper bound for the surface redshift of a compact spin-fluid object can be written as
\begin{equation}
  z\leq \frac{2}{1-3\kappa^2S^2(R)/\left<\rho _{\rm eff}\right>} \, .
\end{equation}
Hence, values of the redshift greater than two may be an observational indicator of the presence of torsion effects in compact bodies.

While the red shift bound (\ref{13}) relates to astrophysical objects, an alternative application of the formalism developed in the present paper relates to the physics of fundamental particles -- in particular, to alternative mathematical models of the strong interaction. One such model is based on the assumption that tensor fields may play an important role in the physical description of strong interactions. This approach is called ``strong gravity'' theory, and was initially proposed and developed in~\cite{strong}. (For alternative approaches to the geometrization of strong interactions see~\cite{strong_alt}.)

\subsection{Spin-generalized strong gravity}

Mathematically, strong gravity is formulated as a two-tensor theory of both the strong and ``ordinary" gravitational interactions, in which equations formally analogous to the Einstein field equations govern the behavior of the strong tensor field. The difference between the ordinary gravitational interaction and the strong gravity interaction results from the different numerical values of the coupling parameters, i.e. $\kappa _f \simeq 1$ GeV$^{-1}$ for the strong interaction versus $k_g \simeq 10^{-19}$ GeV$^{-1}$ for the Newtonian gravitational coupling~\cite{strong}. (Here, we follow the notation used in~\cite{strong_alt}.)

Tentatively, we apply a similar logic to Einstein-Cartan theory, replacing $\kappa \equiv k_g \simeq 10^{-19}$ GeV$^{-1}$ with $\kappa _f \simeq 1$ GeV$^{-1}$ in the field equations, and the corresponding Buchdahl-type inequalities, in order to construct a ``spin-generalized strong gravity" theory. In principle, this should be capable of describing certain aspects of realistic strong physics -- namely, gauge singlet interactions, including effective models of confinement -- to particles with spin.  Strictly, such a generalization of strong gravity theory is in fact {\it necessary} to describe {\it baryons} (not just mesons) but, despite some early investigations \cite{G1},  and, somewhat surprisingly, it has not thus far been fully explored in the literature.

Thus, we assume that the minimum mass bound given by Eq.~(\ref{68}) is valid in the spin-generalized strong gravity theory, with $G\rightarrow G_f=6.67\times 10^{30}$ cm$^3$/s$^2$ g and $\kappa ^2\rightarrow \kappa _f^2=8\pi G_f/c^4$,
respectively. We then obtain the following mass-radius-spin relation for strongly interacting elementary particles:
\begin{equation}
M_{\rm eff} \geq \frac{1}{4}\frac{c^4}{G_f}\kappa _f^4S^2(R)R^3 = (4\pi)^2\frac{G_f}{c^4}S^2(R)R^3 \, .
\end{equation}

For an elementary particle with spin $\hbar/2$ the spin density is given by
\begin{equation}  \label{79}
S^2=\left(\frac{3\hbar}{8\pi R^3}\right)^2=1.582\times 10^{22}\times \left(\frac{R}{1\;\mathrm{fermi}}\right)^{-6}\;\mathrm{g^2/s^2\;cm^2} \, ,
\end{equation}
which gives for the minimum mass bound the expression
\begin{equation} \label{spin_mass_bound}
M_{\rm eff}\geq 2.058\times 10^{-26}\times \left(\frac{R}{1\;\mathrm{fermi}}\right)^{-3}\;\mathrm{g} \, .
\end{equation}
For $R \simeq (l_{\rm Pl}^2l_{\rm dS})^{1/3} \simeq r_{e} = e^2/(m_{e}c^2) \simeq 1$ fermi, as suggested by Eq.~(\ref{me}), $M_{\rm eff}|_{\rm min} \simeq m_{e} \simeq \alpha_{e}(m_{\rm Pl}^2m_{\rm dS})^{1/3} $, as {\it also} suggested by Eq.~(\ref{me}). Hence, the two bounds are self-consistent. In fact, by adopting for $R$ the value $R=r_e=2.81$ fm we can reproduce almost exactly the mass of the electron as $M_{\rm eff}\left(r_e\right)=9.27\times 10^{-28}$ g $\simeq =m_e$. (The exact value of the electron mass is $m_e=9.11\times 10^{-28}$ g.) The same result can be achieved by assuming a particle radius of about $R=1$ fm, and by slightly modifying the value of $G_f$ to $G_f=0.30\times 10^{30}$ cm$^3$/ g s$^2$.

However, we note that (strictly), the absolute lower for the mass-radius ratio of a realistic baryon, ``living" in a dark energy Universe, should be derived from the generalized Buchdahl inequality including $\Lambda >0$, $S^2>0$ and $Q^2>0$, respectively. This lies beyond the scope of the present paper and must be left for future work. That said, we also note the following points:

\begin{enumerate}

\item It is reasonable to assume that the overall spin density of the Universe has a negligible effect on the position of the asymptotic de Sitter horizon $r_{\rm H}(t_0) \simeq l_{\rm dS} = \sqrt{3/\Lambda}$, so that the DE-UP (\ref{gen}) remains valid independently of Eq.~(\ref{68*}).

\item It is reasonable also to neglect the role of cosmological constant in Eqs.~(\ref{charged_min}) and (\ref{68*}) so that the Bekenstein bound $R \gtrsim Q^2/(Mc^2)$ and the strong gravity spin bound (\ref{68}) remain valid independently of each other, and of the DE-UP (\ref{gen}).

\item Since Eq.~(\ref{spin_mass_bound}) comes from evaluating the spin bound (\ref{68}) for the spin of an elementary fermion (\ref{79}), points 1 and 2 imply that the DE-UP, the Bekenstein bound, and the strong gravity spin-bound (\ref{spin_mass_bound}) all hold independently of each other, at least approximately.

\item This, in turn, implies that Eqs.~(\ref{me}) and (\ref{spin_mass_bound}) hold independently.

\end{enumerate}

It is therefore intriguing, and highly suggestive of a fundamental link between gravity, including dark energy, spin, including torsion, and both the strong and electro-weak interactions, that combining the DE-UP with the Bekenstein bound (i.e. combining $\Lambda$ with $\alpha_e$) yields the same mass limit as combining strong gravity with the spin of an elementary fermion (i.e. combining $S^2 \propto \hbar$ with $\kappa_f^2 \propto \alpha_s \simeq 1$). Furthermore, the resulting mass scale is observed in nature, and corresponds to the mass of the electron, $m_e \simeq 10^{-28}$ g. Though, clearly, the electron does not feel the strong force, these results suggest a link between spin ($S^2 \propto \hbar$) and the strong force ($\kappa_f^2 \propto \alpha_s \simeq 1$) and between the electromagnetic force ($\alpha_e = e^2/(\hbar c) \simeq 1/137$) and dark energy ($\rho_{\Lambda} = \Lambda c^2/(8\pi G)$), respectively. Ultimately, this suggests a link between $\alpha_s$, $\alpha_e$ and $\rho_{\Lambda}$, {\it all} of which play a role in determining the mass of elementary particles. At present, the exact nature of the mechanism(s) by which the masses of fundamental particles may be generated from the 4 fundamental forces of nature remains obscure, though investigations along the lines of the analysis presented herein, and in the recent series of papers~\cite{Burikham:2015sro,Burikham:2017bkn,Lake:2017ync,B1,Lake2017} may provide fruitful avenues for future research.

Using the expression (\ref{79}) for the spin density for an elementary particle, the minimum mass bound in strong gravity in the presence of (strong) torsion can be reformulated as
\begin{equation}\label{mimpli}
M_{\rm eff}\geq \frac{9 G_f h^2}{4 c^4 R^3} \, .
\end{equation}
The existence of the Compton wavelength for massive quantum particles implies a minimum localization scale (at low energies) of $\lambda_{C} = \hbar /(Mc)$ for a particle of mass $M$. On the other hand, in general relativity, a ``particle" of mass $M$ cannot be localized to within an region greater than the associated Schwarzschild radius $r_S = 2GM/c^2$. If the particle size is less than this distance, no signal from $r < r_S$ can reach the outside world and gravitational self-trapping occurs.  Setting $\lambda_{C}(M) = r_S(M)$ yields $M \simeq m_{\rm Pl}$, so that the Planck mass marks the boundary between the black hole and elementary particle regimes \cite{BHUP}. In string gravity theory, the equivalent condition yields the ``strong gravity Planck mass"
\begin{equation}  \label{82}
m_{\rm sPl} = \sqrt{\frac{\hbar c}{2G_f}}.
\end{equation}

As noted in \cite{Burikham:2017bkn}, for $\kappa_f \simeq 1 \ \rm GeV^{-1}$, $m_{\rm sPl}$ is of the order of the nucleon mass. More specifically, setting $m_{\rm sPl}=m_n$, where $m_n=1.674\times 10^{-24}$ g is the neutron mass, Eq.~(\ref{82}) gives $G_f\approx 6\times 10^{30}$ cm$^3$/g\;s$^2$ = $10^{38}\;G$, where $G$ is the Newtonian gravitational constant, which is exactly the value of $G_f$ postulated in
the strong gravity theory~\cite{strong}. Hence, using this representation for $G_f$, we obtain for the minimum mass bound in strong gravity the expression
\begin{equation}
M_{\rm eff}\geq \frac{9 \hbar ^3}{8 c^3 m_n^2 R^3}=\frac{9}{8}m_n\frac{\lambda _C^{(n)}}{R^3}=\frac{9M_{\rm min}^3}{8m_n^2},
\end{equation}
where where $\lambda _C^{(n)}=\hbar/m_nc$ is the Compton wavelength of the neutron, and $M_{min}$ is given by
\begin{equation}
M_{\rm min}=\frac{\hbar }{cR}.
\end{equation}

\section{Discussions and final remarks}\label{sect7}

In the present paper, we have considered upper and lower mass-radius ratio bounds for compact spin-fluid spheres in Einstein-Cartan theory, in the presence of a dark energy density generated by a cosmological constant, and of a dark fluid satisfying a linear equation of state with coefficient $0 < w_0 \leq 1$. For simplicity, we assumed throughout that the ordinary thermodynamic density and pressure of the matter fluid also satisfies a linear barotropic equation of state. 
In our analysis, we derived explicit bounds for the mass-radius ratio $2GM/c^2R$ as a function of the spin density of the compact object $S^2$, and of the cosmological constant $\Lambda$ or the generalized dark energy parameters. As our physical model for the spin, we adopted the Weyssenhoff fluid, which represents an unpolarized material (``fluid") consisting of microscopic particles with randomly orientated intrinsic (quantum) spins.

The effects of the spin degree of freedom are extremely important for the cosmological evolution of the very early Universe and, in the spin-fluid dominated epoch, the energy density of the spin-fluid scales as $(1+z)^6$, where $z$ is the cosmological redshift \cite{T3,T4}. However, in order to obtain a description consistent with observational data, the contribution of the spin-fluid cannot dominate over the standard radiation term before the onset of BBN, i.e., before $z\approx 10^8$. Nonetheless, by imposing BBN and CMB constraints, a limit of $\Omega _{s,0}=-0.012$ for the density parameter of a spin fluid is still possible \cite{T3} at the $1\sigma$ level. Though worthy of further study, in the present work we have considered only the effects of the spin-fluid on compact astrophysical objects, in which the torsion gives just a small contribution to the matter energy density and pressure, and have not attempted to analyze the interesting case of the spin-fluid dominated cosmological epoch. Hence, the possibilities of the survival, inside high density stars, of the remnants of the initial torsion determined by spin-fluids, and of avoiding the the Big Bang singularity through torsion contributions to the gravitational field, remain consistent with present day cosmological observations.

In contrast to standard general relativity, we have not found universal limits -- which are independent of both $S^2$ and $\Lambda$ -- for the mass-radius ratio of compact spin-fluid objects. In particular, we found that the spin density plays a key role in determining the minimum mass-radius ratio of a stable, compact, neutral spin-fluid sphere, which we identify with a simgle elementary particle by setting $S^2 \sim \hbar/R^3$. Crucially, we found that such a limit exists for $S^2 > 0$, even in the absence of dark energy ($\Lambda =0$).

However, while the lower bound on the mass-radius ratio may be applicable to elementary particles, the upper bound is of relevance to astrophysical objects. Our analysis shows that the surface red shift of such objects is strongly modified due to the presence of spin, which affects both the energy density and pressure distribution inside the fluid sphere, as well as by the presence of non-vanishing spin density at the vacuum boundary. In general, the mass-radius ratio limits depend on the value of the surface spin density, so that different physical models of the spin could lead to very different upper and lower bounds.

A general feature of the parameters which characterize the physical properties of spin-fluid compact objects in Einstein-Cartan theory is that their absolute limiting values depend on both $\Lambda$ and $S^2$. These include the minimum/maximum mass-radius ratios, and the surface red shift of the system. Tentatively, we have extended our results to the field of the elementary particles via the strong gravity approach initiated in~\cite{strong}, which has been proposed as an effective geometric description of strong interactions. By rescaling the Newtonian gravitational constant $G$ to its strong gravity analogue $G_f \simeq 10^{30}\;G$, we obtained a mass-spin-radius relation for minimum-mass particles in spin-generalized strong gravity theory, $M_{{\rm eff}}\propto G_fS^2R^3$. Evaluated numerically, this is of the same order of magnitude as the mass of the electron. On the other hand, this result also implies the existence of a {\it minimum mass density}, given by  $\rho _{{\rm eff}}=M_{{\rm eff}}/R^3\propto S^2$, which is fully determined by the spin density of the object. Hence, at least at the level of fermionic elementary particles, both mass and mass density appear as manifestations of the essentially quantum property of the existence of intrinsic rotation (spin), which has no classical analogue.

However, in applying the spin-fluid model to elementary particles, we note an intrinsic drawback of our analysis. In this case, we take the spin-fluid description at ``face value" as a model of nuclear matter, rather than as a continuum approximation as in the astrophysical case. Although, in principle, this is theoretically viable, we note that for particles the spin {\it does} have an overall polarisation, i.e. spins are either ``up" or ``down". Whilst, on purely dimensional grounds, we may expect the same or quantitatively similar results will hold, even when the polarisation is explicitly accounted for, it is worth pointing out that our existing model does not explicitly capture this (physical) feature of the elementary constituents of matter.

The possible role of the Einstein-Cartan theory in the physics of elementary particles has also been recently emphasized in~\cite{impli}, where it was proposed that, by using a non-linear extension of the Dirac equation known
as the Hehl-Datta equation, obtained within the Einstein-Cartan-Sciama-Kibble generalization of general relativity, one can solve two of the major fundamental problems in theoretical physics~\cite{impli}:  why no elementary fermionic particles exist in the mass range between the electroweak scale and the Planck scale, and what is the nature of the energy counterbalancing the divergencies of the electrostatic and strong force energies of point-like charged fermions near the Planck scale? By using an S-matrix approach, as well as some semiclassical considerations, an equation giving the radius $r_x$ of an elementary particle of mass $m_x$ can be derived in the form $m_xc^2=e^2/r_x-\left(G/r_x^3\right)\left(\hbar /2c\right)^2$, which for $m_x=m_e$ correctly reproduces the electron radius. On the other hand, in our approach based on the Einstein-Cartan formulation of strong gravity, describing strong interactions, particle masses are naturally generated at the electroweak energy scale, due to their explicit dependence on the quantum mechanical spin density, whose numerical value is fixed by the fundamental laws of quantum mechanics. In fact, the second term in the mass equation for $m_x$ in~\cite{impli} is (almost) the same as the minimum mass given by Eq.~(\ref{mimpli}), written in Newtonian gravity, but with an important sign difference.

At the other end of the scale, an important result in theoretical astrophysics is the existence of a maximum mass for stable, compact, astrophysical objects like white dwarfs and neutron stars. This limiting mass was found by Chandrasekhar and Landau and is known as the Chandrasekhar mass $M_{\rm Ch}$~\cite{Sh}. It is given by
\begin{align*}
  M_{\rm Ch} = \left[\left(\hbar c/G\right)m_B^{-4/3}\right]^{3/2},
\end{align*}
where $m_B$ is the mass of an individual particle (for example, an electron or baryon in the case of white dwarfs or neutron stars, respectively), where a large number of such particles give the main contribution to the mass of the object. For $m_B=m_e$, the Chandrasekhar mass is of the order of $M_{Ch}\approx 1.4\;M_{\odot}$. Hence, it exceeds by many orders of magnitude the mass of any elementary particle.
Moreover, the Chandrasekhar mass is a universal limit, depending only on the fundamental constants of nature, and does not contain the radius of the object. However, in the present paper, we obtained both upper and lower bounds on the mass-radius ratio for compact objects that cannot be represented in the Chandrasekhar form.

In conclusion, the methods developed in the present analysis provide theoretical tools that could aid the experimental detection of the presence of torsion in the natural world, on both astrophysical and elementary particle scales.

\acknowledgments

   We would like to thank the anonymous referee for comments and suggestions that helped us to improve our manuscript.  The work of BH is partly supported by COST Action CA15117 (Cosmology and Astrophysics Network for Theoretical Advances and Training Actions), which is part of COST (European Cooperation in Science and Technology). TH would like to thank the Yat Sen School of the Sun Yat Sen University in Guangzhou, P. R. China, for the kind hospitality offered during the preparation of this work.  ML is supported by a Naresuan University Research Fund individual research grant.


\end{document}